\documentclass[prd,twocolumn,floatfix,footinbib,bibnotes,longbibliography]{revtex4-1}

\pdfoutput=1

\usepackage[colorlinks=true,citecolor=blue,linkcolor=magenta,hypertexnames=false]{hyperref}
\usepackage{amsmath}
\usepackage{graphicx}
\usepackage{dsfont}
\usepackage{changepage}
\usepackage{fancyhdr}
\usepackage{amsthm, amssymb}
\usepackage{float}
\usepackage{array}
\usepackage[usenames,dvipsnames]{color}
\usepackage{slashed}
\usepackage{subfigure}

\def \be{\begin{equation}}
\def \ee{\end{equation}}
\def \bea{\begin{eqnarray}}
\def \eea{\end{eqnarray}}

\begin{document}

\title{Quantized electrochemical transport in Weyl semimetals}

\author{R. Flores-Calder\'{o}n}
\email{rafael.flores@correo.nucleares.unam.mx}
\affiliation{Instituto de Ciencias Nucleares, Universidad Nacional Aut\'{o}noma de M\'{e}xico, 04510 Ciudad de M\'{e}xico, M\'{e}xico}

\author{A. Mart\'{i}n-Ruiz}
\email{alberto.martin@nucleares.unam.mx}
\affiliation{Instituto de Ciencias Nucleares, Universidad Nacional Aut\'{o}noma de M\'{e}xico, 04510 Ciudad de M\'{e}xico, M\'{e}xico}

\begin{abstract}
{

We show that under the effect of an external electric field and a gradient of chemical potential, a topological electric current can be induced in Weyl semimetals without inversion and mirror symmetries. We derive analytic expressions for the nonlinear conductivity tensor and show that it is nearly quantized for small tilting when the Fermi levels are close to the Weyl nodes. When the van Hove point is much larger than the largest Fermi level, the band structure is described by two linearly dispersing Weyl fermions with opposite chirality. In this case, the electrochemical response is fully quantized in terms of fundamental constants and the scattering time, and it can be used to measure directly the topological charge of Weyl points. We show that the electrochemical chiral current may be derived from an electromagnetic action similar to axion electrodynamics, where the position-dependent chiral Fermi level plays the role of the axion field. This posits our results as a direct consequence of the chiral anomaly.a

 }
\end{abstract}
\maketitle

\section{Introduction}

Topological semimetal is a new phase of matter which is characterized by a nontrivial electronic structure topology, making it distinct from an ordinary metal \cite{Armitage, Felser, Burkov}. Topological nontriviality of a semimetal is characterized by momentum space invariants defined on the Fermi surface, rather than in the whole Brillouin zone as in topological insulators. When the Fermi surface is close to a linear crossing of two nondegenerate bands, the low-energy quasiparticles are relativistic Weyl fermions and the contact point is known as Weyl node \cite{Murakami, Burkov2}. Such node is protected from becoming gapped because it carries a monopole source of Berry curvature, and its charge is a topological invariant: the chirality \cite{Xiao}.

A remarkable spectroscopic manifestation of the nontrivial electronic structure topology of WSMs is the existence of topologically protected surface states, which are called Fermi arcs \cite{Wan}, discovered primarily through observation in angle-resolved photoemission spectroscopy \cite{WSM-exp, WSM-exp2, WSM-exp3}. WSMs also display various distinctive transport and optical properties, such as the anomalous Hall effect in a TR broken phase \cite{Steiner, Burkov3}, the chiral magnetic effect \cite{Kharzeev-CME, Landsteiner-CME} and natural optical activity \cite{Tewari-OptAct, Moore-OptAct} in inversion broken WSMs. One of the most fascinating features of WSMs is the chiral anomaly, i.e. the anomalous nonconservation of fermions with a given chirality in the presence of parallel electric and magnetic fields. Consequences of the chiral anomaly in WSMs  are a negative longitudinal magnetoresistence \cite{Spivak-NMR, Burkov-NMR, Li-NMR} and the planar Hall effect \cite{Tewari-PHE, Burkov-PHE, Tewari-PHE2}.

The search for quantized physical observables in condensed matter systems has attracted great attention in the last decades. Outstandingly, the quantum Hall effect in two-dimensional systems and the half-integer Hall effect at topological insulator surfaces are direct consequences of the quantization of the Berry phase \cite{Haldane, Qi}. Quantized responses have also been identified in metallic systems, such as the optical conductivity and optical transmittance in graphene \cite{Gusynin, Falkovsky, Kuzmenko}, and the circular photogalvanic effect (CPGE) in WSMs and three-dimensional Rashba materials \cite{Moore-CPGE}.  In the absence of any scattering mechanism, the CPGE current grows unboundedly with a quantized rate. However, disorder introduce a finite scattering rate $1/ \tau$ and the linear growth of current can only be observed for times $t < \tau$. In the limit $t \gg \tau$ the current saturates to the previously predicted quantized value multiplied by the scattering time. On the other hand, the anomaly induced chiral magnetic effect (CME) was proposed as a potentially quantized linear response in WSMs as well \cite{Kharzeev, Zyuzin, Tewari}. However, we now know that the CME is determined by orbital moments  rather than the chiral anomaly \cite{Tewari2} and its magnitude depends on a nonuniversal parameter: the energy offset between the Weyl nodes.

In this paper we show that in Weyl semimetals without inversion and mirror symmetries, the Berry curvature related current response induced by an electric field $\vec{E}$ and a gradient of a local chemical potential $\mu (\vec{r} \, )$, becomes a truly quantized response. Obviously, the analytic function $\mu (\vec{r} \, )$ is completely determined by its value and the value of its derivative at any point of the sample. Experimentally, the local chemical potential can be controlled by doping or having inhomogeneous impurities in the sample \cite{Rodionov1, Rodionov2}. Unlike their high-energy counterparts, Weyl cones in condensed matter systems can also be anisotropic, tilted, and they can be connected due to the bending of the bands. Here, we first consider a simple low-energy continuum model with band-bending and tilting. For small tilts, we find that the current response is anisotropic, with nearly quantized components multiplied by the disorder-induced finite scattering time $\tau$. If the chemical potential is brought near to each of the Weyl nodes, e.g. by balancing the concentrations of donors and acceptors impurities, the model reduces to that of two linearly-dispersing massless fermions with opposite chiralities, and the electrochemical current becomes a truly quantized response times $\tau$, independent of the energy splitting between Weyl points. Such chiral current is regarded as being a direct consequence of the chiral anomaly of the WSM phase, derivable from an electromagnetic action similar to axion electrodynamics, where the position-dependent chiral Fermi level plays the role of the axion field.

This paper is organized as follows. In Sec. \ref{Model} we introduce the model we shall consider: a two-band Hamiltonian for a tilted and band-bended Weyl semimetal phase. In Sec. \ref{KineticEqSect} we solve the Boltzmann transport equation in the presence of external electric field and gradient of chemical potential. This allow us to establish a general formula for the nonlinear conductivity tensor as a function of the local chemical potential. The main results of this paper, a nearly quantized topological response in the presence of band-bending and a truly quantized response in the chiral limit, are presented in Sec. \ref{ElectrochemicalWSM}. In Sec. \ref{Conclusions} we summarize our results and give further concluding remarks.

\section{The Model} \label{Model}

The effective low-energy continuum Hamiltonian we shall consider is given by
\begin{align}
H _{\vec{k}} = \hbar \, \vec{t} \cdot \vec{k} \, \sigma _{0} + \hbar v _{F} \, \vec{\sigma} \cdot \vec{K} , \label{Hamiltonian}
\end{align}
where $\vec{t}$ is the tilt velocity and $v _{F}$ is the Fermi velocity of electrons. The Pauli matrices $\vec{\sigma}$ represents an effective orbital basis, not necessarily the spin degree of freedom. Since unbounded linear dispersion model is not realistic in solid state systems, we introduce band-bending effects through the modified crystal momenta $\vec{K} = k _{x} \vec{e} _{x} + k _{y} \vec{e} _{y} + K _{z} \vec{e} _{z}$, where $K _{z} = (k _{z} ^{2} - b ^{2}) / 2b$. The Hamiltonian (\ref{Hamiltonian}) preserves particle-hole symmetry, but breaks both time-reversal and parity symmetries.

The eigenstates of the Hamiltonian (\ref{Hamiltonian}) are calculated as
\begin{align}
\left| \right. \! \psi _{s \vec{k}} \! \left. \right> = \frac{1}{\sqrt{2K (K + s K _{z})}} \left( \begin{array}{c} K _{z} + sK \\ k _{x} + i k _{y} \end{array} \right) , \label{Eigenstates}
\end{align}
with the eigenenergies
\begin{align}
E _{s \vec{k}} = \hbar \, \vec{t} \cdot \vec{k} + s \hbar v _{F} K , \label{Eigenenergies}
\end{align}
where $K = \sqrt{k _{\rho} ^{2} + K _{z} ^{2}}$, $k _{\rho} ^{2} = k _{x} ^{2} + k _{y} ^{2}$, and $s = \pm 1$ is the band index. 

\begin{figure}
\centering
\subfigure[]{\includegraphics[scale=0.232]{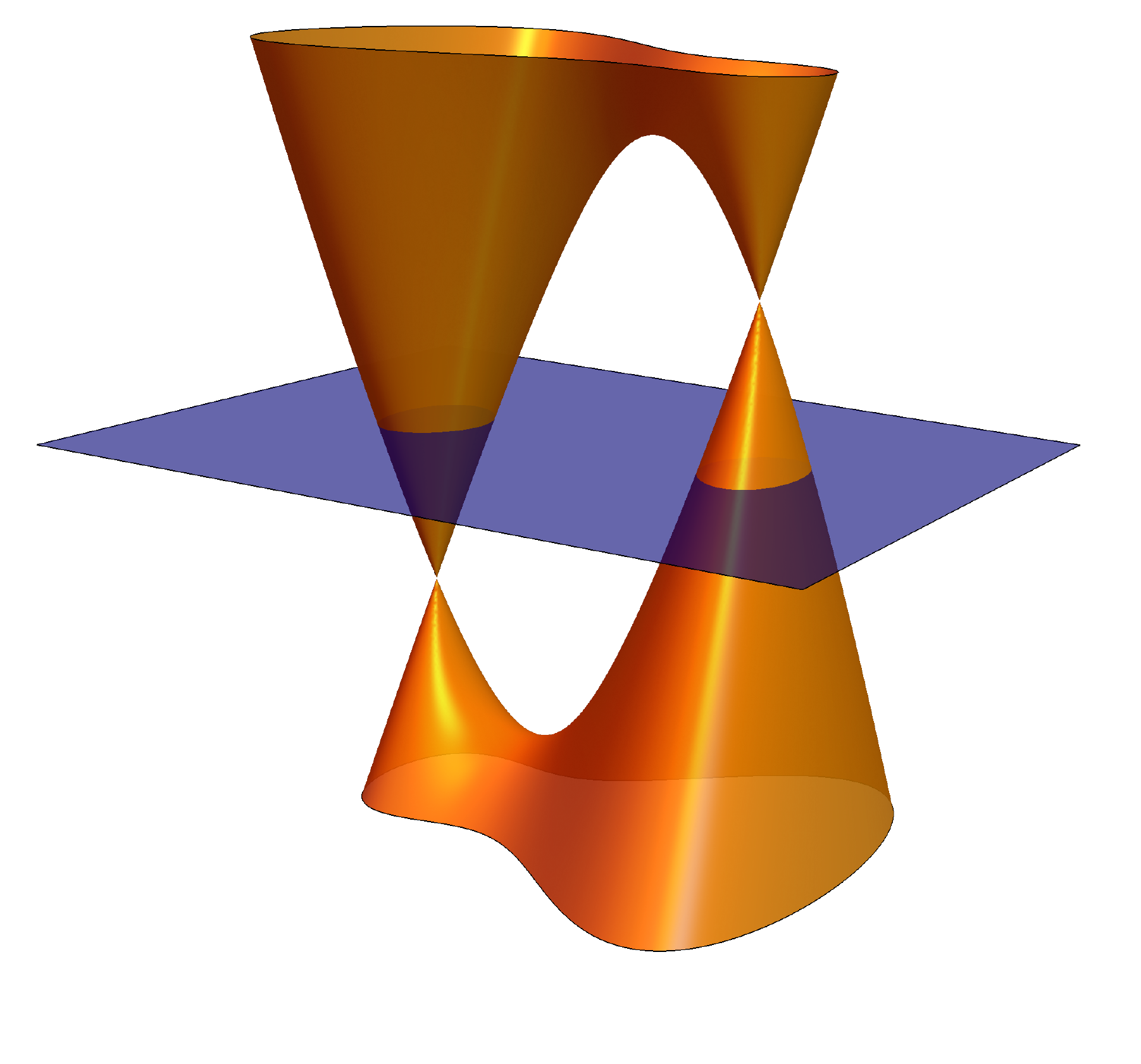}}
\subfigure[]{\includegraphics[scale=0.232]{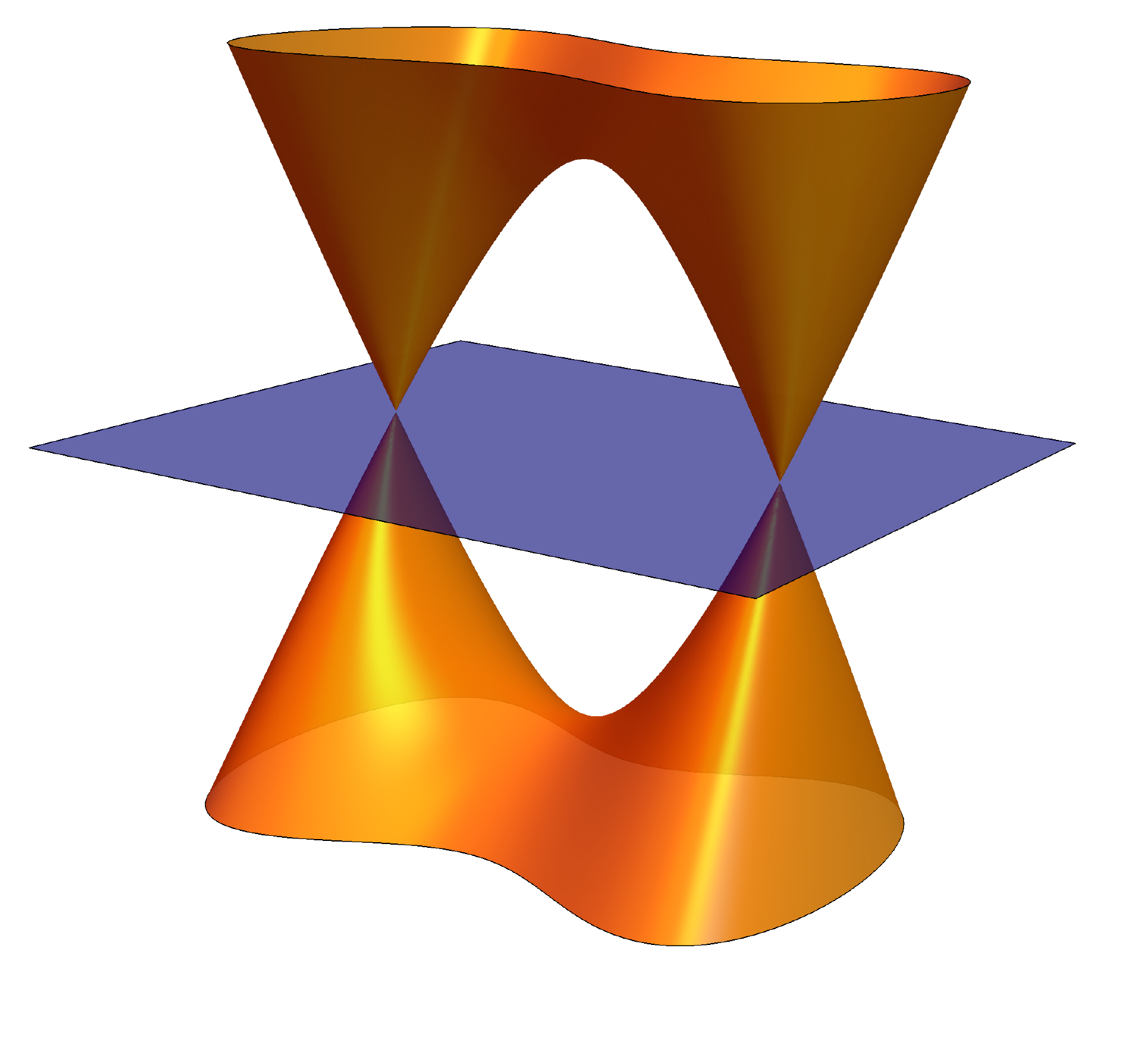}}
\caption{Schematic illustration of the band structure for a Weyl semimetal (a) tilted along the $k _{z}$ axis and (b) with vanishing tilting vector. The zero-energy plane is shown in blue for reference.}
\label{DispersionsPlot}
\end{figure}

In Fig. \ref{DispersionsPlot} we plot the dispersion of the energy bands in different situations, and we show the zero-energy plane in blue for reference. The spectrum consists of two bands touching at two Weyl points located at $\vec{b} = \pm b \vec{e} _{z}$. Figure \ref{DispersionsPlot}(a) shows the band structure for a WSM tilted along the $k _{z}$ axis. Due to the tilting, the band-touching points are separated in energy by an amount $2 \hbar t b$. This reflects the broken parity symmetry of the model.  Figure \ref{DispersionsPlot}(b) corresponds to the energy dispersion for a WSM with broken time-reversal symmetry. In this case, the band-touching points lies on the zero-energy surface. In the following, we consider the tilt only in the $k _{z}$ direction, because it produces electron-hole pockets, leading to interesting transport phenomena. In-plane tilting does not produce interesting effects unless it is so extreme that the system is in the type-II semimetallic phase.

Explicit expressions for the group velocity $\vec{v} _{s \vec{k}}$, the Berry curvature $\vec{\Omega} _{s \vec{k}} $ and the orbital magnetic moment $\vec{m} _{s \vec{k}}$, will be necessary  to compute the transport coefficients in Sec. \ref{ElectrochemicalWSM}.  So, we proceed to derive them. The group velocity if found to be 
\begin{align}
\vec{v} _{s \vec{k}} = \frac{1}{\hbar} \vec{\nabla} _{\vec{k}} E _{s \vec{k}} = \vec{t} + v _{F} \frac{s}{K} \left[ k _{\rho} \, \vec{e} _{\rho} + K _{z} \frac{\partial K _{z}}{\partial k _{z}} \vec{e} _{z} \right] . \label{Velocity}
\end{align}
The Berry curvature and orbital magnetization of the energy bands are defined in terms of the the Bloch eigenfunctions (\ref{Eigenstates}) as
\begin{align}
\vec{\Omega} _{s \vec{k}} &= i \left< \right. \! \vec{\nabla} _{\vec{k}} \, \psi _{s \vec{k}} \! \left. \right| \times \left| \right. \! \vec{\nabla} _{\vec{k}} \, \psi _{s \vec{k}} \! \left. \right> , \label{BerryCurvDef} \\ \vec{m} _{s \vec{k}} &= - i \frac{e}{2 \hbar}  \left< \right. \! \vec{\nabla} _{\vec{k}} \, \psi _{s \vec{k}} \! \left. \right| \times \left( H _{\vec{k}} - E _{s \vec{k}} \right) \left| \right. \! \vec{\nabla} _{\vec{k}} \,  \psi _{s \vec{k}} \! \left. \right> , \label{OrbMagDef}
\end{align}
respectively. They are intrinsic properties of the band structure because they only depend on the wave functions. Further, they do not become affected by the tilt as the tilt enters the Hamiltonian with an identity matrix. Using the Bloch eigenfunctions (\ref{Eigenstates}) and the definition (\ref{BerryCurvDef}) it is straightforward to compute the Berry curvature. The result is 
\begin{align}
\vec{\Omega} _{s \vec{k}} = - \frac{s}{2 K ^{3}} \left[ k _{\rho} \frac{\partial K _{z}}{\partial k _{z}} \vec{e}_{\rho} + K _{z} \vec{e} _{z} \right]  . \label{BerryCurvature}
\end{align}
The orbital magnetization is found to be proportional to the Berry curvature. Indeed one can spot the identity $\vec{m} _{s \vec{k}} = s e v _{F} K \, \vec{\Omega} _{s \vec{k}}$.

\section{Anomalous electrochemical transport in kinetic theory} \label{KineticEqSect}

We will now investigate the nonlinear electrochemical current in the quasiclassical Boltzmann formalism. We are interested in the situation with external homogeneous and static electric field $\vec{E}$, and gradient of chemical potential $\vec{\nabla} _{\vec{r}} \, \mu$. 

It has been established that the Berry phase of Bloch states has a profound effect on transport driven by mechanical forces \cite{Niu1, Niu2}. In the presence of Berry curvature $\vec{\Omega} _{s \vec{k}}$, the semiclassical equation of motion for an electron in the band $s$ takes the following form
\begin{align}
\dot{\vec{r}} _{s} &= \vec{v} _{s \vec{k}}  - \dot{\vec{k}}  \times \vec{\Omega} _{s \vec{k}} \label{EqMot.r} , 
\end{align}
where $\vec{v} _{s \vec{k}} = \frac{1}{\hbar} \vec{\nabla} _{\vec{k}} E _{s \vec{k}}$ is the band velocity and $\vec{k}$ is the crystal momentum. In the presence of an electric field $\vec{E}$ we also have the usual Lorentz force relation  $\dot{\vec{k}} = - (e/ \hbar )\vec{E}$. The first term in Eq. (\ref{EqMot.r}) is the familiar relation between semiclassical velocity $\dot{\vec{r}} _{s}$ and the band energy dispersion $E _{s \vec{k}}$. The second term gives rise to anomalous transport perpendicular to the applied electric field. Evaluation of the Hall current from this term reproduces the Karplus-Luttinger formula for the anomalous Hall conductivity \cite{Luttinger}.

Berry phase also manifests in transport driven by a statistical force, such as the gradient of temperature or chemical potential. Unlike mechanical forces which can be described by perturbation to the Hamiltonian for the carriers, statistical forces manifest at the macroscopic level and makes sense only through the statistical distribution of carriers \cite{Niu3}. In the conventional Boltzmann transport theory, one considers a statistical distribution function $f _{s} (t, \vec{k} , \vec{r} \, )$ of carriers in the phase space of position and crystal momentum. The quasiparticle distribution function satisfies the Boltzmann equation \cite{Ashcroft}
\begin{align}
\left( \frac{\partial}{\partial t} + \dot{\vec{r}} _{s} \cdot \vec{\nabla} _{\vec{r}} + \dot{\vec{k}} \cdot \vec{\nabla} _{\vec{k}} \right) \! f _{s} ( \vec{k} , \vec{r} \, ) = I _{\mbox{\scriptsize coll}} [ f _{s} ( \vec{k} , \vec{r} \, ) ] , \label{BoltzmanEq}
\end{align}
where on the right side $I _{\mbox{\scriptsize coll}} [ f _{s} ] $ is the collision integral whose form depends on the details of the collision process. In this paper we use the relaxation-time approximation for the collision integral,
\begin{align}
I _{\mbox{\scriptsize coll}} [ f _{s} (\vec{k} , \vec{r} \, ) ] = - \frac{f _{s} (\vec{k} , \vec{r} \, ) - f _{s} ^{(0)} (\vec{k} , \vec{r} \, )}{\tau} , \label{RelaxationTimeApp}
\end{align}
where $\tau$ is the scattering time of quasiparticles, which for simplicity we take as independent of momentum; and $f _{s} ^{(0)} (\vec{k} , \vec{r} \, )$ is the equilibrium \textit{local} distribution function defined by local temperature $T(\vec{r} \, )$ and local chemical potential $\mu (\vec{r} \, )$. The assumption that the main effect of the scattering processes is the restoration of local thermodynamic equilibrium on the time scale given by $\tau$ defines the relaxation-time approximation of the Boltzmann equation \cite{Ashcroft}.

In what follows, we consider the steady-state Boltzmann equation in the relaxation-time approximation:
\begin{align}
\vec{v} _{s \vec{k}} \cdot \vec{\nabla} _{\vec{r}} \, f _{s} + \frac{e}{\hbar} \vec{E} \cdot ( \vec{\Omega} \times \vec{\nabla} _{\vec{r}} - \vec{\nabla} _{\vec{k}} ) \, f _{s} = - \frac{f _{s} - f _{s} ^{(0)}}{\tau} , \label{BoltzmanEq2}
\end{align}
where we have omitted the dependence of the distribution function on $\vec{r}$ and $\vec{k}$ for simplicity. Since we are interested in the transport properties of an electron fluid, $f _{s} ^{(0)}$ corresponds to the Fermi-Dirac distribution in the absence of any external mechanical force. The solution of Eq. (\ref{BoltzmanEq2}) will be used to compute the nonequilibrium quasiparticle current density \cite{Niu4}
\begin{align}
\vec{J} &= - e \sum _{s = \pm 1} \int \frac{d ^{3} \vec{k}}{(2 \pi) ^{3}} \; \left( \dot{\vec{r}} _{s} - \frac{1}{e} \vec{\nabla} _{\vec{r}} \times \vec{m} _{s \vec{k}} \right) f _{s} (\vec{k},\vec{r} \, ) , \label{current}
\end{align}
where the second term is a contribution from the magnetization current. In this expression $\vec{m} _{s \vec{k}} $ is the orbital magnetic moment (\ref{OrbMagDef}), which generically describes the rotation of a wave packet around its center of mass \cite{Niu1, Niu2}.

As usual, one can recursively solve the Boltzmann equation (\ref{BoltzmanEq2}) assuming that $f _{s} = f _{s} ^{(0)} + f _{s} ^{(1)} + f _{s} ^{(2)}$, where $f _{s} ^{(1)} \sim \mathcal{O} ( E _{i} ) + \mathcal{O} ( \partial _{i} \mu)$ and  $f _{s} ^{(2)} \sim \mathcal{O} (E _{i} \, \partial _{j} \mu)$ contain the linear and the nonlinear terms, respectively. One finds
\begin{align}
f _{s} ^{(1)} &= \tau \, \vec{v} _{s \vec{k}} \cdot ( - e \vec{E} + \vec{\nabla} _{\vec{r}} \, \mu ) \; \frac{\partial f _{s} ^{(0)}}{\partial E _{s \vec{k}}} , \label{First-orderDistribution} \\ f _{s} ^{(2)} &= \frac{e \tau}{\hbar} ( \vec{E} \times \vec{\Omega} _{s \vec{k}} \cdot  \vec{\nabla} _{\vec{r}} \, \mu ) \; \frac{\partial f _{s} ^{(0)}}{\partial E _{s \vec{k}}} . \label{Second-OrderDistribution} 
\end{align} 
There are other terms not listed in $f _{s} ^{(2)}$ which are not related to the Berry curvature. Inserting these results into Eq. (\ref{current}) we find the quasiparticle currrent density. The nonlinear current, proportional to $\vec{E}$ and $\vec{\nabla} _{\vec{r}} \, \mu$, can be written as the sum of three terms:
\begin{align}
\vec{J} & = - e \sum _{s = \pm 1} \int \frac{d ^{3} \vec{k}}{(2 \pi) ^{3}} \, \bigg[ \vec{v} _{s \vec{k}} \, f _{s} ^{(2)} + \frac{e}{\hbar} \vec{E} \times \vec{\Omega} _{s \vec{k}} \, f _{s} ^{ (1)} \notag \\ & \hspace{3.5cm}  + \frac{1}{e}  \vec{m} _{s \vec{k}} \times \vec{\nabla} _{\vec{r}} \, f _{s} ^{(1)} \bigg]   .  \label{SecondOrderCurrent} 
\end{align}
The first term in Eq. (\ref{SecondOrderCurrent}) is the usual convective current, the second term is the anomalous Hall current, and the last term is the magnetization current that originates from the orbital magnetic moment of quasiparticles.

\section{Electrochemical transport in Weyl Semimetals}  \label{ElectrochemicalWSM}

With the aid of the above results, it is straightforward to calculate various anomalous electrochemical responses. Here, we first focus on the Weyl semimetal model of Eq. (\ref{Hamiltonian}). For the present model the orbital magnetization is found to be proportional to the Berry curvature. In fact one can spot the identity $\vec{m} _{s \vec{k}} = s e v _{F} K \, \vec{\Omega} _{s \vec{k}}$. Let us define the three terms appearing in the current of Eq. (\ref{SecondOrderCurrent}) as $\vec{J} \equiv \vec{\mathcal{J}} _{1} + \vec{\mathcal{J}} _{2} + \vec{\mathcal{J}} _{3} $, respectively. A careful algebraic manipulation of the currents in Eq. (\ref{SecondOrderCurrent}) together with the axial symmetry of the problem gives
\begin{align}
\vec{\mathcal{J}} _{1} &= \frac{\tau e ^{2}}{h ^{2}} \left[ I _{\perp} (\mu ) \, ( \vec{\nabla} \mu \times \vec{E} ) _{\perp} + I _{z} (\mu ) \, ( \vec{\nabla} \mu \times \vec{E} ) _{z} \right] , \label{current1} \\ \vec{\mathcal{J}} _{2} &= \frac{\tau e ^{2}}{h ^{2}} \left[ I _{\perp} (\mu ) \, ( \vec{E} \times \vec{\nabla} _{\perp} \mu ) + I _{z} (\mu ) \, ( \vec{E} \times \vec{\nabla} _{z} \mu ) \right] , \label{current2} \\ \vec{\mathcal{J}} _{3} &= \frac{\tau e ^{2}}{h ^{2}} \left[ F _{\perp} (\mu ) \, ( \vec{\nabla} \mu \times \vec{E} _{\perp} ) + F _{z} (\mu ) \, ( \vec{\nabla} \mu \times \vec{E} _{z} ) \right] , \! \label{current3} 
\end{align}
where we have defined the dimensionless functions
\begin{align}
I _{\perp} (\mu ) &= - \frac{\hbar}{2 \pi} \sum _{s = \pm 1} \int d ^{3} \vec{k} \; v _{s \vec{k}} ^{x} \, \Omega _{s \vec{k}} ^{x} \, \frac{\partial f _{s} ^{(0)}}{\partial E _{s \vec{k}}} , \label{Iperp} \\ I _{z} (\mu ) &= - \frac{\hbar}{2 \pi} \sum _{s = \pm 1} \int d ^{3} \vec{k} \; v _{s \vec{k}} ^{z} \, \Omega _{s \vec{k}} ^{z} \, \frac{\partial f _{s} ^{(0)}}{\partial E _{s \vec{k}}} , \label{Iz}
\end{align}
and
\begin{align}
F _{\perp} (\mu ) &= \frac{v _{F} \hbar ^{2}}{2 \pi} \sum _{s = \pm 1} \int d ^{3} \vec{k} \; s \, K \, v _{s \vec{k}} ^{x} \, \Omega _{s \vec{k}} ^{x} \, \frac{\partial ^{2} f _{s} ^{(0)}}{\partial E _{s \vec{k}} ^{2}} , \label{Fperp} \\ F _{z} (\mu ) &=  \frac{v _{F} \hbar ^{2}}{2 \pi} \sum _{s = \pm 1} \int d ^{3} \vec{k} \; s \, K \, v _{s \vec{k}} ^{z} \, \Omega _{s \vec{k}} ^{z} \, \frac{\partial ^{2} f _{s} ^{(0)}}{\partial E _{s \vec{k}} ^{2}} . \label{Fz}
\end{align}
Now we have to evaluate these dimensionless integrals. Of course, they depend strongly on where the chemical potential is located with respect to the band touching points and the van-Hove points. Details of technical computations are relegated to the Appendix \ref{AppA-Computation}. 

\begin{figure}
\includegraphics[scale=0.6]{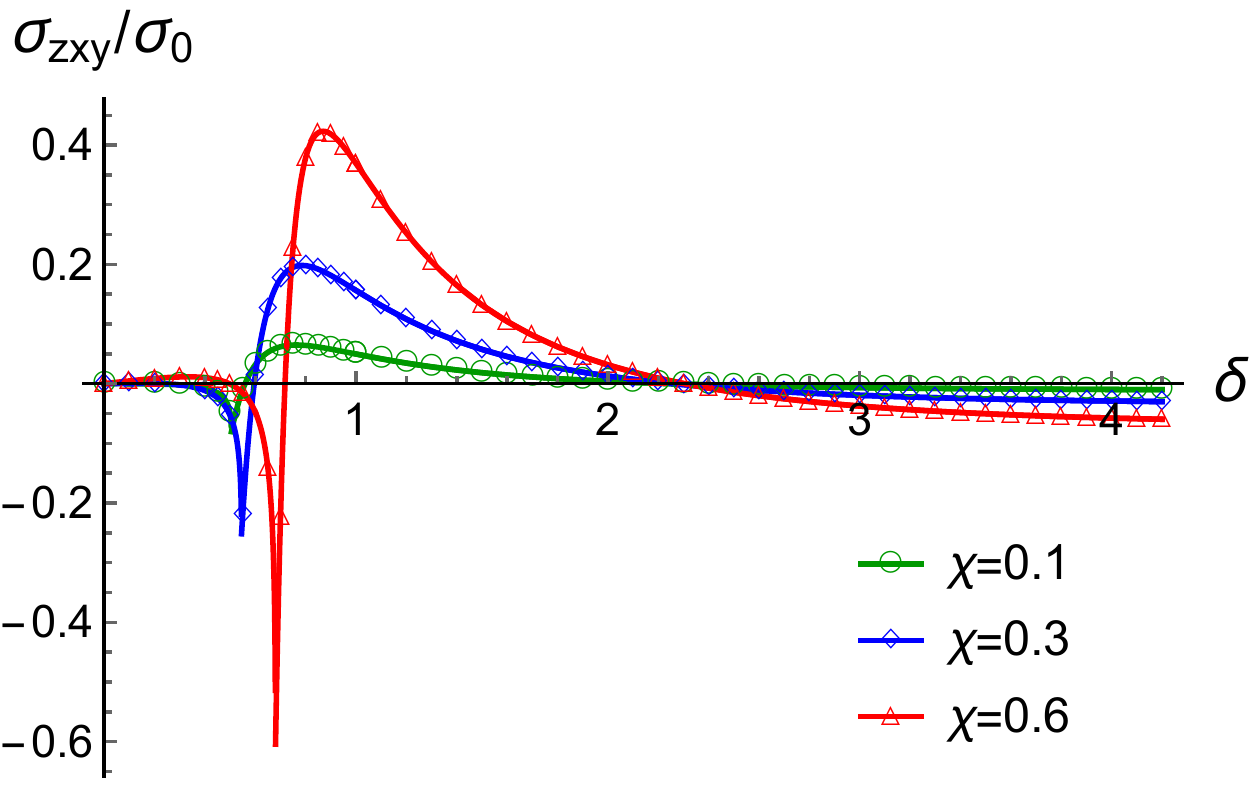}
\caption{Behaviour of the conductivity $\sigma _{zxy}$ (in units of conductance quantum $\sigma _{0} = e ^{2} \tau / h ^{2}$) as a function of the dimensionless Fermi level $\delta = \mu / ( \hbar v _{F} b )$ and different values of $\chi = t / v _{F}$. }
\label{Sigmazxy}
\end{figure}

In general, the electrochemical current (\ref{SecondOrderCurrent}) can be expressed as $J _{i} = \sigma _{ijk} (\mu ) E _{j} \partial _{k} \mu$ (latin indices span the cartesian components $\{\ \!\! x,y,z \}\ \!\!$), where the nonlinear conductivity tensor $\sigma _{ijk}$ is a function of the chemical potential $\mu$, the tilt velocity $t$, and the Weyl nodes separation $b$. Indeed, it can be expressed as a linear combination of the functions $I _{\alpha}$ and $F _{\alpha}$, where $\alpha = \perp , z$. It is worth mentioning that the electrochemical current is completely determined by the value of the chemical potential and the value of its derivative at any point of the sample. A fully analytic expression for $\sigma _{ijk} (\mu )$ (for a given Fermi level $\mu$) do not provide too much physical insight. Instead we plot a particular component of the nonlinear conductivity tensor in Fig. \ref{Sigmazxy}. 

From now on we introduce a notation for the dimensionless Fermi level $\delta = \mu / ( \hbar v _{F} b )$ and the dimensionless tilting strength $\chi = t / v _{F}$.  In Fig. \ref{Sigmazxy} we observe an interesting behaviour when $\mu$ is close to the van Hove point [with dimensionless energy $\delta _{ \mbox{\scriptsize vH}} = (1 + \chi ^{2}) / 2 \,$] or the Weyl nodes separation is too small. In such cases, where the notion of chirality is lost, $\sigma _{zxy} (\mu )$ is naturally peaked exactly at the vH point (due to the singularity in the density of states). As $\mu$ increases above $\epsilon _{\mbox{\scriptsize vH}}$, the conductivity function exhibits an extrema (which is direct manifestation of the band-bending, so it is absent in the chiral model) and then asymptotically tends to zero.

\subsection{Nearly quantized case}

As we know, chirally imbalanced Weyl semimetals exhibit interesting transport phenomena. For example, the chiral magnetic effect, which is the generation of a current by an applied magnetic field, and the quantized circular photogalvanic effect \cite{Moore-CPGE}, which is the production of a dc current by circularly polarized light incident on a surface material. Despite their prediction as potentially quantized responses, none of them have been shown to be truly quantized. In the former case, the CME depends on the energy offset between the Weyl nodes, whilst in the latter case, it was recently shown that interactions (Coulomb and Hubbard) remove the quantization of the chiral photocurrent at Weyl points \cite{Moore-CPGE2}. As we shall see below, in a Weyl semimetal where the nodes lie at different Fermi levels, the electrochemical response becomes a truly quantized response that depends only on fundamental constants. 

\begin{figure*}
\includegraphics[scale=0.35]{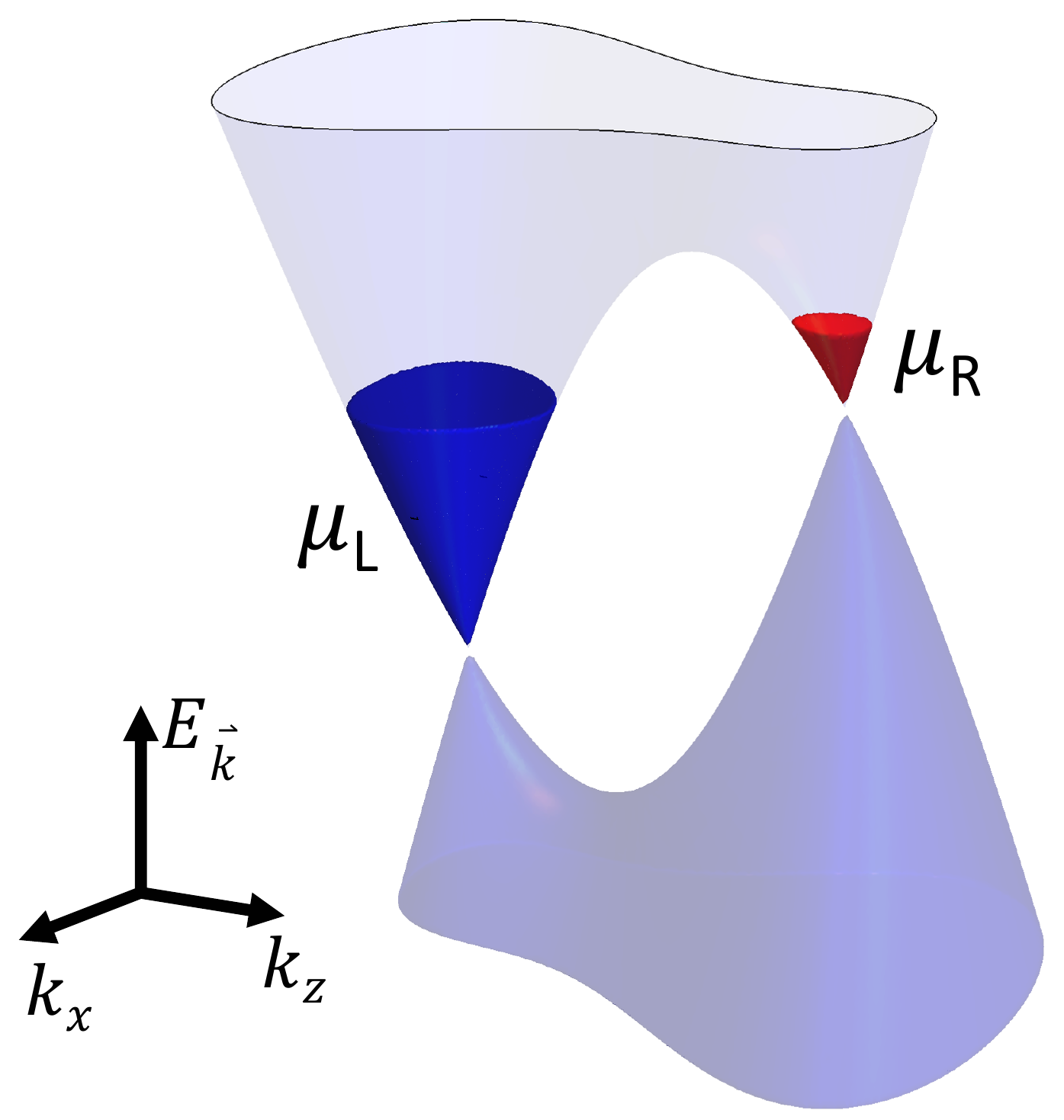} \hspace{2cm}
\includegraphics[scale=0.9]{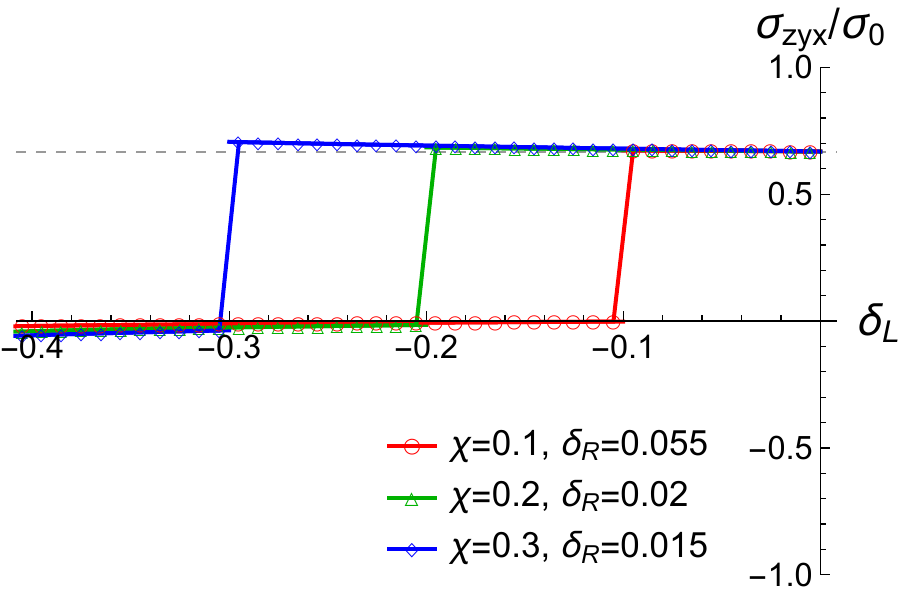}
\caption{Left: Band structure of a Weyl semimetal with Weyl nodes ($L$ and $R$). The Fermi levels $\mu _{L}$ and $\mu _{R}$ at the nodes are measured from the energy of the band touching points. Right: Conductivity $\sigma _{zyx}$ (in units of conductance quantum $\sigma _{0} = e ^{2} \tau / h ^{2}$) as a function of the dimensionless left-Fermi level $\delta _{L} = \mu _{L} / ( \hbar v _{F} b )$. For left-Fermi levels $\mu _{L}$ between the Weyl node energies, the electrochemical response is quantized to $2/3$ (see main text). }
\label{SigmazxyQuantized}
\end{figure*}

In a realistic Weyl semimetal the Fermi level is determined by doping by donor and acceptor impurities. When only one type of impurity dominates, the Fermi sea is homogeneous near each node, whose depth $\mu _{i}$ significantly exceeds the characteristic fluctuations of the disorder potential arising from the randomly-localized impurities. Interestingly, by balancing the concentrations of donors and acceptors, it is possible to bring the Fermi level close to both nodal points \cite{Rodionov1, Rodionov2}. This is precisely the situation we shall consider in the following. 

Let $\mu _{L}$ and $\mu _{R}$ be the Fermi level of the left- and right- node as measured from the band crossing points, respectively. See the left panel of Fig. \ref{SigmazxyQuantized}. When the Fermi levels are such that $\vert \mu _{L} \vert = \mu _{R}$, the electrochemical current is zero. This is similar to what happens with the CME and the CPGE, which also vanish for a WSM in equilibrium. In the case $\vert \mu _{L} \vert \neq \mu _{R}$,  the nonlinear conductivity tensor $\sigma _{ijk}$ (in units of conductance quantum $\sigma _{0} = e ^{2} \tau / h ^{2}$) is nearly quantized over some range of the chiral Fermi level $( \mu _{R} - \vert \mu _{L} \vert ) / 2$. In Appendix \ref{NearlyQ-App} we present the details of the calculations. In the right panel of Fig. \ref{SigmazxyQuantized} we plot $\sigma _{zyx}$ as a function of the (dimensionless) left-Fermi level $\delta _{L} = \mu _{L} / (\hbar v _{F} b)$ and different values of $\delta _{R} = \mu _{R} / (\hbar v _{F} b)$ and $\chi = t / v _{F}< 1$. We observe that $\sigma _{zyx} / \sigma _{0}$ is nearly quantized to $2/3$ as long as $\delta _{L} > - \chi $, which means that the left-Fermi level $\mu _{L}$ lies between the Weyl node energies. This yields to the nearly quantized nonlinear conductivity tensor
\begin{align}
\sigma _{ijk} = - 2 \sigma _{0} \left( \delta _{kt} \epsilon _{ijt} + \frac{1}{3} \delta _{it} \epsilon _{tjk} \right) \Theta (\delta _{L} + \chi ) , \label{NearlyQConductivity}
\end{align}
where $t$ refers to the direction of the tilt and $\Theta$ is the Heaviside function.  Note that although the magnitude of the components of the conductivity tensor does not depend on the tilt, the size of the plateau is precisely controlled by the magnitude and sign of the tilt, as dictated by the step function in Eq. (\ref{NearlyQConductivity}). All in all, the nearly quantized electrochemical current becomes
\begin{align}
\vec{J} = 2 \sigma _{0} \left[ \vec{\nabla} _{t} \, \mu \times \vec{E} + \frac{1}{3} (\vec{\nabla} \mu \times \vec{E}) _{t} \, \vec{e} _{t} \right] . \label{NearlyQCurrent}
\end{align} 
Note that, similar to what happens with the quantized circular photogalvanic effect in which the current saturates to a quantized coefficient multiplied by the relaxation time, our electrochemical current is expressed by a nearly quantized coefficient ($\sim e ^{2} / h ^{2}$) multiplied by the disorder-induced finite scattering rate $\tau$. However, there is a fundamental difference: while the CPGE is quantized in whole sample, the electrochemical response (\ref{NearlyQCurrent}) is quantized locally, i.e. point by point along the sample. This means that the spatial profile of the current (\ref{NearlyQCurrent}) is the same as the gradient profile of the chemical potential.

\subsection{Fully quantized case}

A fully quantized coefficient arises when the chemical potentials are very close to the nodal points. In such case, band-bending effects are negligible, and the model Hamiltonian (\ref{Hamiltonian}) reduces to that of two linearly-dispersing massless fermions with opposite chiralities. Another possibility for the quantized response to be observed is in Weyl semimetals with inversion and all mirror symmetries broken, such as in enantiomorphic crystals, allowing for the nodes to occur at different energies.

To see this, let us consider that the nodes $L$ and $R$ have, respectively, negative and positive chiralities, i.e. they are described by the Hamiltonian 
\begin{align}
H _{\chi \vec{k}} = \Delta _{\chi} + \chi \hbar v _{F} \; \vec{\sigma} \cdot (\vec{k} - \chi \vec{b}) , \label{HamiltonianChiral-Supp}
\end{align}
where $\chi = \pm 1$ is the chirality eigenvalue for a given Weyl node and the vector $\chi \vec{b}$ localizes the node with chirality $\chi$ from the origin at $\vec{k} = \vec{0}$. Also, $\Delta _{\chi}$ are constants defining the energy of the Weyl node with chirality $\chi$. It is not necessarily induced by tilting, but in general by broking mirror symmetry. For definiteness we consider the Fermi level $\mu _{\chi}$ crossing the corresponding conduction band. In this case, the energy dispersion, velocity and Berry curvature are written as
\begin{align}
E _{\chi \vec{k}} &=  \Delta _{\chi} + \hbar v _{F} \sqrt{k _{\rho} ^{2} + (k _{z} - \chi b) ^{2}} , \\ \vec{v} _{\chi \vec{k}} &= v _{F} \frac{k _{\rho} \vec{e} _{\rho} + (k _{z} - \chi b) \vec{e} _{z}}{\sqrt{k _{\rho} ^{2} + (k _{z} - \chi b) ^{2}}} , \label{VelChiral} \\ \vec{\Omega} _{ \chi \vec{k}} &= - \chi \frac{\vec{k} - \chi \vec{b}}{2 \vert \vec{k} - \chi \vec{b} \vert ^{3}} . \label{BerryCChiral}
\end{align}
for the conduction band.

Now we apply the semiclassical formula, Eq. (\ref{SecondOrderCurrent}), for the electrochemical current response $\vec{J}$ to the WSM model of Eq. (\ref{HamiltonianChiral-Supp}). Inserting these results into Eq. (\ref{SecondOrderCurrent}) we find that it can be written as the sum of three terms, $\vec{J} =\vec{\mathfrak{J}} _{1} + \vec{\mathfrak{J}} _{2} + \vec{\mathfrak{J}} _{3}$, where
\begin{align}
\vec{\mathfrak{J}} _{1} &= - \vec{\mathfrak{J}} _{2} = - \sum _{\chi = \pm 1}  \frac{e ^{2} \tau}{8 \pi ^{3} \hbar ^{2}} \, (  \vec{\nabla} _{\vec{r}} \,  \mu _{\chi} \times \vec{E} ) \; I _{\chi} , \notag \\ \vec{\mathfrak{J}} _{3} &= - \sum _{\chi = \pm 1}  \frac{e ^{2} \tau}{8 \pi ^{3} \hbar ^{2}} \, ( \vec{E} \times  \vec{\nabla} _{\vec{r}} \, \mu _{\chi}  ) \; F _{\chi} , \label{ChiralCurrents}
\end{align}
where we have defined the dimensionless integrals 
\begin{align}
I _{\chi} &= \frac{\hbar}{3} \int ( \vec{v} _{\chi \vec{k}}  \cdot \vec{\Omega} _{\chi \vec{k}} ) \;  \frac{\partial f _{\chi} ^{(0)} }{\partial E _{\chi \vec{k}}} d ^{3} \vec{k} , \label{I-integral} \\ F _{\chi} &= - \frac{\hbar}{3}  \int E _{\chi \vec{k}} \;  ( \vec{v} _{\chi \vec{k}} \cdot \vec{\Omega} _{\chi \vec{k}} ) \frac{\partial ^{2} f _{\chi} ^{(0)}}{\partial E _{\chi \vec{k}} ^{2}} d ^{3} \vec{k} . \label{F-integral}
\end{align}
Here, $f _{\chi} ^{(0)}$ is the Fermi-Dirac distribution for the Weyl fermion of chirality $\chi$, and we have assumed that  the relaxation time is the same for right- and left-handed fermions. In Appendix \ref{FullyQ-App} we present a detailed solution of these integrals. The final result is $I _{\chi}  = F _{\chi} = (2 \pi / 3 ) \chi$. All in all, inserting the function $F _{\chi} $ into Eq. (\ref{ChiralCurrents}) we obtain a simple expression for the electrochemical current
\begin{align}
\vec{J}  = \frac{2}{3} \sigma _{0} \,  \vec{\nabla} _{\vec{r}} \, \mu _{5} \times  \vec{E}  , \label{ChiralCurrent}
\end{align}
where $\mu _{5} = (\mu _{R} - \mu _{L} ) / 2$ is the chiral chemical potential and  $\sigma _{0} = e ^{2} \tau / h ^{2}$ is the conductance quantum. This expression is similar to that for the anomalous Hall effect, $\vec{J} _{b} = \frac{e ^{2}}{\pi h} \vec{b} \times \vec{E}$, which depends only on the distance between the Weyl nodes in momentum space \cite{Steiner, Burkov3}. Remarkably $\vec{J} _{b}$ and $\vec{J}$ may be regarded as being direct consequences of the chiral anomaly of the WSM phase, as we shall discuss in the rest of this section.

\subsection{Chiral anomaly induced transport}

To see that the electrochemical response induced by the chiral anomaly does in fact have observable consequences, it is useful to note that the current (\ref{ChiralCurrent}) may be obtained from the functional variation of the following action for the electromagnetic field:
\begin{align}
S _{5} = \frac{4}{3} \sigma _{0} \int d ^{4} x \, \mu _{5} (\vec{r} \, ) \vec{E} \cdot \vec{B} , \label{S5Action}
\end{align}
i.e. $\vec{J} = \delta S _{5} / \delta \vec{A}$. This action is similar to that of axion electrodynamics \cite{Wilczek}, with the position-dependent chiral chemical potential $\mu _{5} (\vec{r} \, )$ playing the role of the axion angle. Equation (\ref{S5Action}) can be easily derived by computing the polarization and magnetization of the sample due to a spatially varying chiral chemical potential. To this end, we first determine the induced charge density. This can be done from the expression for the electrochemical current $\vec{J} = \frac{2}{3} \sigma _{0} \,  \vec{\nabla} _{\vec{r}} \, \mu _{5} \times  \vec{E} $ together with the continuity equation $\vec{\nabla} \cdot \vec{J} + \partial _{t} \rho = 0$. These give rise to a charge density $\rho = - \frac{2}{3} \sigma _{0} \vec{B} \cdot \vec{\nabla} \mu _{5}$, which is also derivable upon variation of the action (\ref{S5Action}) with respect to the scalar potential $\phi$, i.e. $\rho = \delta S _{5} / \delta \phi$. 

Apart from the nonelectrochemical response of the material, these charge $\rho$ and current $\vec{J}$ densities reveal new contributions to the polarization $\vec{P}$ and magnetization $\vec{M}$, which are related by $  \rho = - \vec{\nabla} \cdot \vec{P} $ and $\vec{J} = \vec{\nabla} \times \vec{M} + \frac{\partial \vec{P}}{\partial t}$, respectively. Combining these results we obtain the following expressions for the local polarization and magnetization vectors:
\begin{align}
\vec{P} (\vec{r} \, ) = \frac{2}{3} \sigma _{0} \mu _{5} (\vec{r} \, ) \, \vec{B} , \qquad \vec{M} (\vec{r} \, ) = \frac{2}{3} \sigma _{0}  \mu _{5} (\vec{r} \, ) \, \vec{E}  ,  \label{Pol-Mag}
\end{align}
which are direct consequences of the nonzero position-dependent chiral chemical potential. These results reveal an interesting Berry-phase induced effect due to the electrochemical response: a \textit{topological magnetoelectricity}. A magnetoelectric effect is defined as a magnetization induced by an electric field, or alternatively, a charge polarization induced by a magnetic field. This is precisely what Eq. (\ref{Pol-Mag}) describes. In general, the part of the electromagnetic action arising from bound currents is of the form $\int d ^{4} x \, ( \vec{E} \cdot \vec{P} + \vec{B} \cdot \vec{M} )$. Inserting the polarization $\vec{P}$ and magnetization $\vec{M}$ given by Eq. (\ref{Pol-Mag}) we easily establish Eq. (\ref{S5Action}). Clearly, this result resembles the topological field theory which describes the anomalous Hall effect:
\begin{align}
S _{\theta} &= \frac{\alpha }{4 \pi ^{2}} \int d ^{4} x \,   \theta (\vec{r} ,t ) \,  \vec{E} \cdot \vec{B} . \label{LagrangianTheta}
\end{align}
Here, $\theta  (\vec{r} ,t )  = 2 ( \vec{b} \cdot \vec{r} - b _{0} t )$ is an axion field, where $2 \vec{b}$ is the separation between the Weyl nodes in momentum space and $2b _{0} = \sum _{\chi = \pm 1} \Delta _{\chi} $ is the separation between the nodes in energy. 

The above results, (\ref{S5Action}) and (\ref{LagrangianTheta}), may be regarded as being a consequence of the chiral anomaly. We observe that $a _{\mu} \equiv \frac{4}{3} ( \tau  / \hbar ) \, \partial _{\mu} \mu _{5}$ are components of a chiral gauge field, which couples linearly to the chiral current 
\begin{align}
J ^{\mu} _{5} = \frac{e ^{2}}{4 \pi ^{2} \hbar} \epsilon ^{\mu \nu \alpha \beta} A _{\nu} \partial _{\alpha} A _{\beta} \label{ChiralCurrent5}
\end{align}
as $a _{\mu} J ^{\mu} _{5}$, just as ordinary electromagnetic gauge field  couples linearly to the electric current as $A _{\mu} J ^{\mu}$.The functional derivative of the action (\ref{S5Action}) with respect to the gauge field $a _{\mu}$ produces the chiral current (\ref{ChiralCurrent5}), and taking the divergence we find 
\begin{align}
\partial _{\mu} J ^{\mu} _{5} = \frac{e ^{2}}{2 \pi ^{2} h} \vec{E} \cdot \vec{B} ,
\end{align}
which we recognize as the chiral anomaly. Importantly, the action (\ref{S5Action}) has a topological origin, since it contains gauge fields coupled through the fully antisymmetric tensor and multiplied by a universal coefficient.

\section{Conclusions and outlook}  \label{Conclusions}

In this paper we have analyzed the anomalous electrochemical transport in Weyl semimetals. Tuning the chemical potentials near to the left- and right-node respectively, we found that the topological response is nearly quantized for small tilting. In such case,  i.e. when band-bending effects can be disregarded, the model can be approximated by two chiral fermions with oppossite chiralities, and the resulting electrochemical current $\vec{J}$ is found to be quantized. Analogous quantization phenomena has been predicted to occur in WSMs, such as the chiral magnetic effect and the quantized circular photogalvanic effect; however, none of them have been shown to be truly quantized.  

\begin{figure}
	\centering
	\includegraphics[scale=0.75]{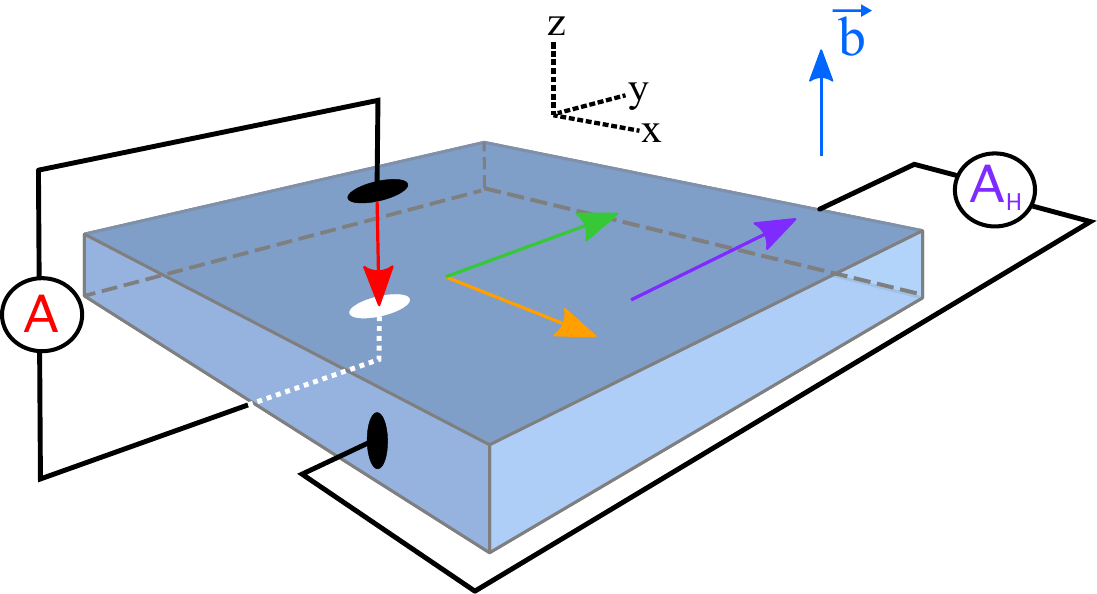}
	\caption{Schematic illustration for the measurement of the electrochemical and anomalous Hall currents in a WSM sample. The direction of the Weyl nodes separation, electric field and gradient of the chiral chemical potential are shown in blue, yellow and green, respectively. Two ammeters, $\text{A}$ and $\text{A} _{H}$, can be used to simultaneously measure the electrochemical current (shown in red) and the anomalous Hall current (shown in purple), respectively.}	\label{Exp-Current-Schematic}
\end{figure}

The importance of our results stem from the fact that the electrochemical current clearly has a topological origin. It can be derived by functionally varying the action $S _{5}$ with respect to the electromagnetic gauge field $\vec{A}$. Also, the nonconservation of chiral charge for Weyl fermions is obtained upon variation of $S _{5}$ with respect to the chiral gauge field $a _{\mu} = \frac{4}{3} ( \tau  / \hbar ) \, \partial _{\mu} \mu _{5}$. One may thus expect the corresponding response to be robust and detail independent, which is always of great interest, and of potential technological importance.

We finally point out that the electrochemical current is an experimentally observable signature of the nontrivial topology of the WSM phase. Figure \ref{Exp-Current-Schematic} schematically illustrates the measurement of the electrochemical current. By choosing the electric field pointing to the $x$-direction, $\vec{E} = E _{0} \vec{e} _{x}$, we observe that a longitudinal chiral chemical potential gradient $\vec{\nabla} \mu _{5} (\vec{r} \, ) = \mu _{5} ^{\prime} (y) \vec{e} _{y}$ produces a local electrochemical current along the $z$-direction, i.e. $J _{z} (y) = \frac{2}{3} \sigma _{0} E _{0} \mu _{5} ^{\prime} (y)$. This current response can be clearly distinguished from the anomalous Hall effect, which in this case is constant and points along $y$-direction, i.e. $J _{b y} = \frac{e ^{2}}{\pi h} b E _{0}$. So, they are to be measured with ammeters placed in different faces of the sample as shown in Fig. \ref{Exp-Current-Schematic}. Observing the nearly quantized electrochemical current requires both the chemical potentials near to the Weyl nodes and an inversion broken WSM. Another possibility is breaking inversion symmetry by shear strain \cite{Cortijo}. The best candidate is perhaps $\text{Sr} \text{Si} _{2}$ \cite{Huang}, since all mirror symmetries are broken, the chemical potential is close to one of the nodes, and the nodes are separated in energy. So, by doping the WSM by donor and acceptor impurities \cite{Rodionov1, Rodionov2}, a chemical potential gradient can be induced and hence provoking the appearance of the electrochemical current.

\acknowledgements 

We thank Eduardo Barrios, Chumin Wang and Luis Urrutia for useful comments and suggestions. A.M.-R. has been partially supported by DGAPA-UNAM Project No. IA101320 and by Project CONACyT (México) No. 428214.

\appendix

\section{Computation of the functions $I _{\alpha} (\mu )$ and $F _{\alpha} (\mu )$} \label{AppA-Computation}

In this section we evaluate explicitly the coefficients $I _{\alpha}$ and $F _{\alpha}$, with $\alpha = \perp , z$, which determine the nonlinear response of the system. Here we will work at zero temperature $T =0$, such that $\partial f _{s} ^{(0)} / \partial E _{s \vec{k}} = - \delta ( \mu - E _{s \vec{k}} )$. This means that the nonlinear transport coefficients are  properties of the Fermi surface.

Let us first consider the functions $I _{\alpha}$, given by Eqs. (\ref{Iperp}) and (\ref{Iz}). Due to the axial symmetry of the Hamiltonian $H _{\vec{k}}$, we shall use cylindrical coordinates to evaluate the required integrals. Substituting the velocity (\ref{Velocity}) and Berry curvature (\ref{BerryCurvature}) into Eqs. (\ref{Iperp}) and (\ref{Iz}) we obtain a generic expression of the form
\begin{align}
I _{\alpha} (\mu ) &= - \frac{\hbar v _{F}}{4 b } \sum _{s = \pm 1} \int _{- \infty} ^{+ \infty} dk _{z} \int _{0} ^{\infty} dk _{\rho} \notag \\ & \hspace{2cm} \times k _{\rho} \, \mathcal{I} _{s \alpha} (k _{\rho} , k _{z})  \delta ( \mu - E _{s \vec{k}} ) , \label{SM-Iperp} 
\end{align}
where
\begin{align}
\mathcal{I} _{s \perp} (k _{\rho} , k _{z}) = \frac{k _{z} k _{\rho} ^{2}}{K ^{4}} , \quad \mathcal{I} _{s z} (k _{\rho} , k _{z}) = \frac{2K _{z}}{K ^{3}} \left[ \frac{s bt}{v _{F}} + \frac{k _{z}  K _{z}}{K} \right] . \label{IntegrandFuncts}
\end{align}
The integral in Eq. (\ref{SM-Iperp}) can be simplified using the following property of the Dirac delta function:
\begin{align}
\delta [f(x)] = \sum _{i} \frac{\delta (x - x _{i})}{\vert f ^{\prime} (x _{i}) \vert} , \label{DiracDeltaProperty}
\end{align}
where $x _{i}$ are the zeros of $f(x)$. In the present case we have $\mu = E _{s \vec{k}}$, wherefrom we obtain the roots
\begin{align}
k _{\rho s \pm} ^{\star} = \pm \sqrt{\left[ \frac{s (\mu - t \hbar k _{z})}{\hbar v _{F}} \right] ^{2} - K _{z} ^{2}} .
\end{align}
Since our integral (\ref{SM-Iperp}) is only for positive values of $k _{\rho}$ (it is a radius in polar coordinates), we are left with only one root, which corresponds to $k _{\rho s} ^{\star} = k _{\rho s +} ^{\star} > 0$. Therefore we can write
\begin{align}
\delta ( \mu - E _{s \vec{k}} ) = \frac{\vert s (\mu - t \hbar k _{z}) \vert}{\hbar ^{2} v _{F} ^{2} k _{\rho s} ^{\star}} \, \delta (k _{\rho} - k _{\rho s} ^{\star} ) \, \Theta (k _{\rho s} ^{\star \,2} ) ,
\end{align}
where the Heaviside function guarantees that the root $k _{\rho s} ^{\star}$ is real valued. We observe that the band index affects the absolute value in the right-hand-side; however, it can be omitted from the root $k _{\rho s} ^{\star}$ since $s ^{2} = 1$. Therefore we henceforth drop the band index in the root, i.e. $k _{\rho} ^{\star} = k _{\rho s} ^{\star}$.

Inserting this result into the integral expression (\ref{SM-Iperp}) and integrating with respect to $k _{\rho}$ we obtain
\begin{align}
I _{\alpha} (\mu ) &= - \frac{1}{4 b \hbar v _{F} } \sum _{s = \pm 1} \int _{- \infty} ^{+ \infty} dk _{z} \; \vert s (\mu - t \hbar k _{z}) \vert \notag \\ & \hspace{3.3cm} \times \mathcal{I} _{s \alpha} (k _{\rho} ^{\star} , k _{z})  \Theta (k _{\rho} ^{\star \,2} ) . \label{SM-Iperp2} 
\end{align}
The step function $\Theta (k _{\rho} ^{\star \,2} )$ gives the integration limits for $k _{z}$, which are determined from the condition $k _{\rho} ^{\star \,2} = 0$. The algebraic equation $k _{\rho} ^{\star \,2} (k _{z} ^{\star}) = 0$ has four solutions:
\begin{align}
k _{z \pm \pm ^{\prime}} ^{\star} (\mu) = b \left( \pm \chi \pm ^{\prime} \sqrt{1 + \chi  ^{2} \mp 2 \delta } \right) ,
\end{align}
where $\chi \equiv t / v _{F}$ is the ratio between the tilt velocity and the Fermi velocity, and $\delta = \mu / \hbar v _{F} b$. In order to understand the physical meaning of these solutions, let's go back to the energy spectrum. A simple analysis reveals that the energy bands (\ref{Eigenenergies}) at $k _{\rho} = 0$ exhibit local extrema in the region $k _{z} \in [-b,b \, ]$ only for $t < v _{F}$. The extremum for the band $s$ is located at $k _{z s} ^{\ast} = s b \chi$, and the extreme value for the energy are the van Hove points: $E _{s \vec{k}} (k _{\rho} = 0 , k _{z} = k _{z s} ^{\ast} ) \equiv E _{s \mbox{\scriptsize vH}} = s \frac{b \hbar v _{F}}{2} (1 + \chi ^{2} )$. In the following discussion we also need the energy of the band touching points: $E _{\pm \mbox{\scriptsize bt}} = \pm \hbar t b $.

\begin{figure*}[t]
\centering
\subfigure[]{\includegraphics[scale=0.18]{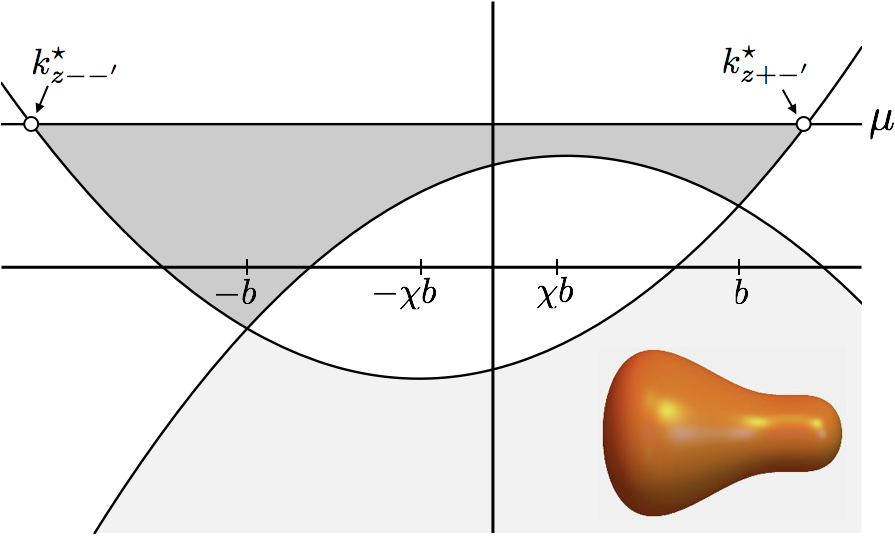}}
\subfigure[]{\includegraphics[scale=0.18]{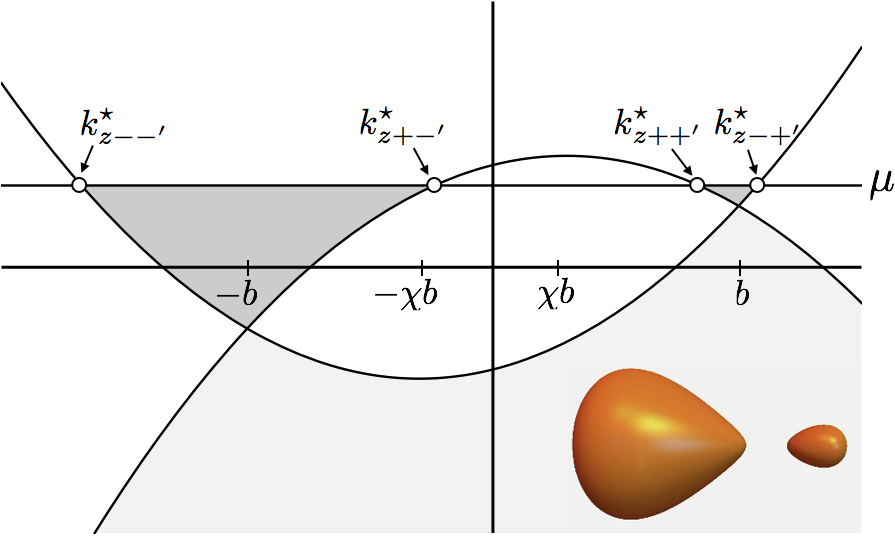}}
\subfigure[]{\includegraphics[scale=0.18]{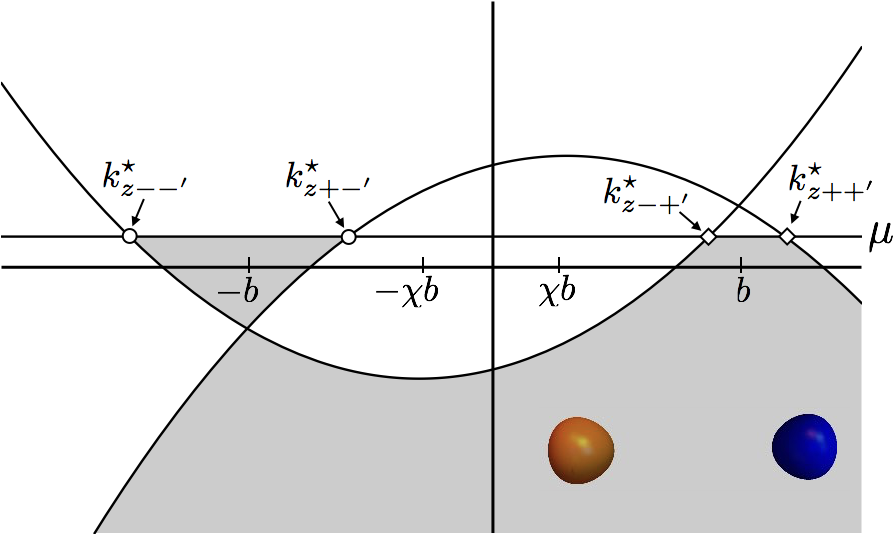}}
\caption{Schematic illustration of the integration regions for the functions $I_\alpha$ and $F_\alpha$ for a given chemical potential $\mu$. The points $k _{z \pm \pm ^{\prime}} ^{*}$ represent the crossings in the conduction or valence bands which act as limits of integration. The insets show the corresponding Fermi surfaces. The orange (blue) lobe corresponds to the conduction (valence) band. }
\label{IntRegions}
\end{figure*}

Based on the above analysis we have, for $\mu > 0$:
\begin{itemize}

\item When the Fermi level lies above the van Hove point $E _{+ \mbox{\scriptsize vH}} $ we have $k _{z + \pm ^{\prime}} ^{\star} \in \mathbb{C}$ and $k _{z - \pm ^{\prime}} ^{\star} \in \mathbb{R}$; i.e. the chemical potential crosses the conduction band twice, and the integration region is $ \boldsymbol{\mathcal{C}} (\mu) = \left\lbrace k _{z} \in \mathbb{R} \; \vert \; k _{z - - ^{\prime}} ^{\star} < k _{z} < k _{z - + ^{\prime}} ^{\star} \right\rbrace$. See Fig. \ref{IntRegions}(a).

\item Now, if the chemical potential is above the band-touching point at $k _{z} = b$ and below the van Hove point $E _{+ \mbox{\scriptsize vH}}$, we have $k _{z \pm \pm ^{\prime}} ^{\star} \in \mathbb{R}$. So, the chemical potential crosses the conduction band four times, and the integration regions are $ \boldsymbol{\mathcal{C}} _{L} (\mu) = \left\lbrace k _{z} \in \mathbb{R} \; \vert \; k _{z - - ^{\prime}} ^{\star} < k _{z} < k _{z + - ^{\prime}} ^{\star} \right\rbrace $ and $ \boldsymbol{\mathcal{C}} _{R} (\mu) = \left\lbrace k _{z} \in \mathbb{R} \; \vert \; k _{z ++ ^{\prime}} ^{\star} < k _{z} < k _{z - + ^{\prime}} ^{\star} \right\rbrace $. See Fig. \ref{IntRegions}(b).

\item If we take the chemical potential below the band touching point and above the zero-energy plane, we have again four solutions. Nevertheless, the chemical potential now crosses both the conduction and valence bands, twice each one, as shown in Fig. \ref{IntRegions}(c). The integration regions now become $ \boldsymbol{\mathcal{C}} _{L} (\mu) = \left\lbrace k _{z} \in \mathbb{R} \; \vert \; k _{z - - ^{\prime}} ^{\star} < k _{z} < k _{z + - ^{\prime}} ^{\star} \right\rbrace $ and $ \boldsymbol{\mathcal{V}} _{R} (\mu) = \left\lbrace k _{z} \in \mathbb{R} \; \vert \; k _{z - + ^{\prime}} ^{\star} < k _{z} < k _{z ++ ^{\prime}} ^{\star} \right\rbrace $.

\end{itemize}

As a consequence of the above analysis for the integration regions we can further simply the integral expression (\ref{SM-Iperp2}) since:
\begin{align}
\vert s (\mu - \hbar t k _{z} ) \vert = \left\lbrace \begin{array}{l} \mu - \hbar t k _{z} , \\[7pt] \hbar t k _{z} - \mu , \end{array} \;\; \begin{array}{l} k _{z} \in \boldsymbol{\mathcal{C}}, \boldsymbol{\mathcal{C}} _{L}, \boldsymbol{\mathcal{C}} _{R}  \\[7pt] k _{z} \in \boldsymbol{\mathcal{V}} _{R}  \end{array} \right. .
\end{align}
Now we are ready to compute the $I _{\alpha}$ coefficients. Explicitly we have

\begin{widetext}
\begin{align}
I _{\alpha} (\mu ) &= - \frac{1}{4b \hbar v _{F}} \Bigg\{\ \!\! \Theta (\mu - E _{ \mbox{\scriptsize vH}} ) \int _{\boldsymbol{\mathcal{C}} (\mu)}  + \,  \Theta (\mu - \hbar b t ) \, \Theta (E _{ \mbox{\scriptsize vH}} - \mu ) \left(  \int _{\boldsymbol{\mathcal{C}} _{L} (\mu)} + \int _{\boldsymbol{\mathcal{C}} _{R} (\mu)} \right) + \Theta (\hbar b t - \mu ) \, \Theta ( \mu ) \notag \\[10pt] &  \hspace{5cm} \times \left(  \int _{\boldsymbol{\mathcal{C}} _{L} (\mu)} - \int _{\boldsymbol{\mathcal{V}} _{R} (\mu)} \right)  \Bigg\}\ \, (\mu - \hbar t k _{z}) \;  \mathcal{I} _{s \alpha} (k _{\rho} ^{\star} , k _{z}) \, dk _{z} .
\end{align}
Considering $\mu > 0$, the first term gives the coefficient for the chemical potential above the van-Hove point [see Fig. \ref{IntRegions}(a)], the second one corresponds to the chemical potential lying between the band touching point and the van-Hove point [see Fig. \ref{IntRegions}(b)], and the third term gives the coefficient below the band touching point [see Fig. \ref{IntRegions}(c)]. Introducing the dimensionless variable $\alpha = k _{z} / b$, the above integral simplifies to
\begin{align}
I _{\perp} (\delta , \chi ) &= - \frac{1}{4} \left\lbrace \Theta (\delta - \delta  _{ \mbox{\scriptsize vH}} ) \int _{\alpha _{--'}} ^{\alpha _{-+'}} + \Theta (\delta ) \Theta ( \delta  _{ \mbox{\scriptsize vH}} - \delta) \left( \int _{\alpha _{--'}} ^{\alpha _{+-'}} + \int _{\alpha _{++'}} ^{\alpha _{-+'}} \right)   \right\rbrace \frac{\alpha}{(\delta - \chi \alpha ) ^{3}} \left[ (\delta - \chi \alpha ) ^{2} - \frac{1}{4} (\alpha ^{2} - 1) ^{2} \right] d \alpha , \\ I _{z} (\delta , \chi ) &= - \frac{1}{4} \left\lbrace \Theta (\delta - \delta  _{ \mbox{\scriptsize vH}} ) \int _{\alpha _{--'}} ^{\alpha _{-+'}} + \Theta (\delta ) \Theta ( \delta  _{ \mbox{\scriptsize vH}} - \delta) \left( \int _{\alpha _{--'}} ^{\alpha _{+-'}} + \int _{\alpha _{++'}} ^{\alpha _{-+'}} \right)   \right\rbrace \frac{\alpha ^{2} - 1}{(\delta - \chi \alpha ) ^{2}} \left[ \chi + \frac{\alpha (\alpha ^{2} - 1)}{2 (\delta - \chi \alpha )} \right] d \alpha , 
\end{align}
where $\alpha _{\pm \pm '} = k _{\pm \pm '} / b = \pm \chi \pm ^{\prime} \sqrt{1 + \chi  ^{2} \mp 2 \delta }$ and $\delta _{ \mbox{\scriptsize vH}} = E _{ \mbox{\scriptsize vH}} / ( \hbar v _{F} b ) = (1 + \chi ^{2}) / 2$. These integrals can be directly evaluated. The full expressions are not illuminating, and hence we do not write them here. Instead, in Figs. \ref{Functions}(a) and \ref{Functions}(b) we plot  the functions $I _{\perp} (\delta , \chi )$ and $I _{z} (\delta , \chi )$, respectively, as a function of $\delta > 0$ and a given value of $\chi < 1$.  The identity $I _{z} (\delta , \chi ) = - 2 I _{\perp} (\delta , \chi )$ can also be directly verified.

\begin{figure*}[ht!]
\centering
\includegraphics[scale=0.7]{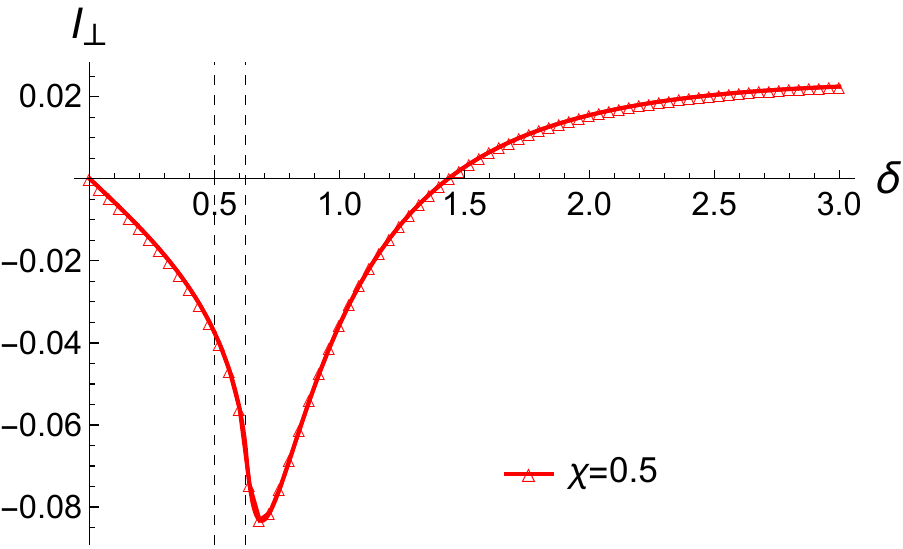}
\hspace{0.1cm}
\includegraphics[scale=0.7]{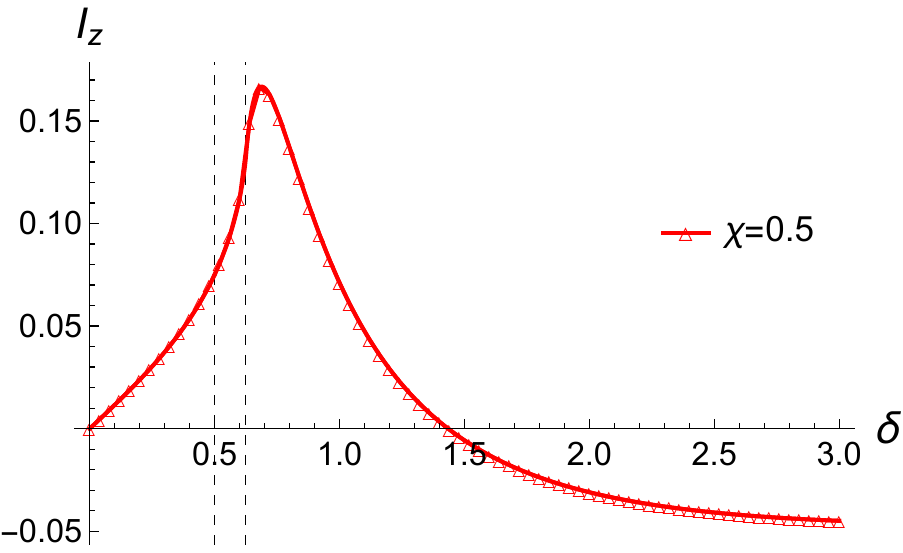} \\ (a) \hspace{8cm} (b) \\
\includegraphics[scale=0.7]{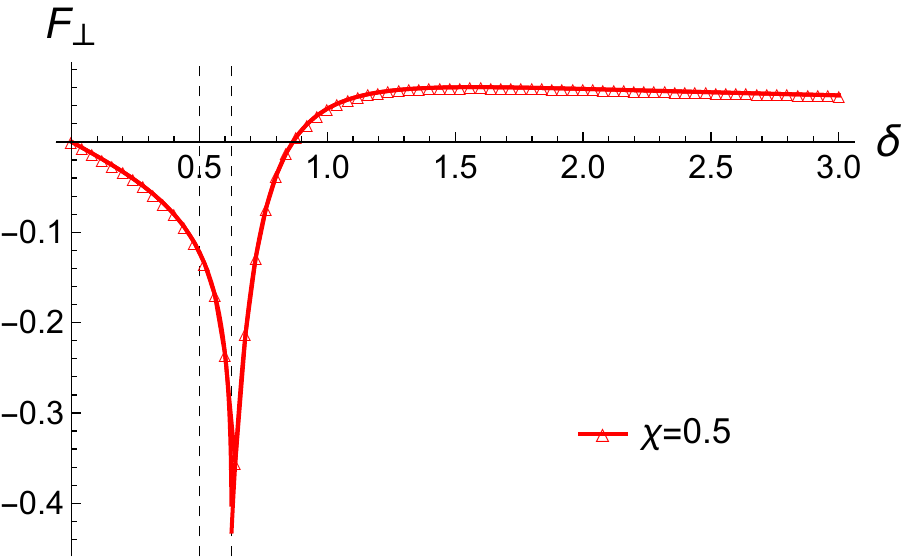}
\hspace{0.1cm}
\includegraphics[scale=0.7]{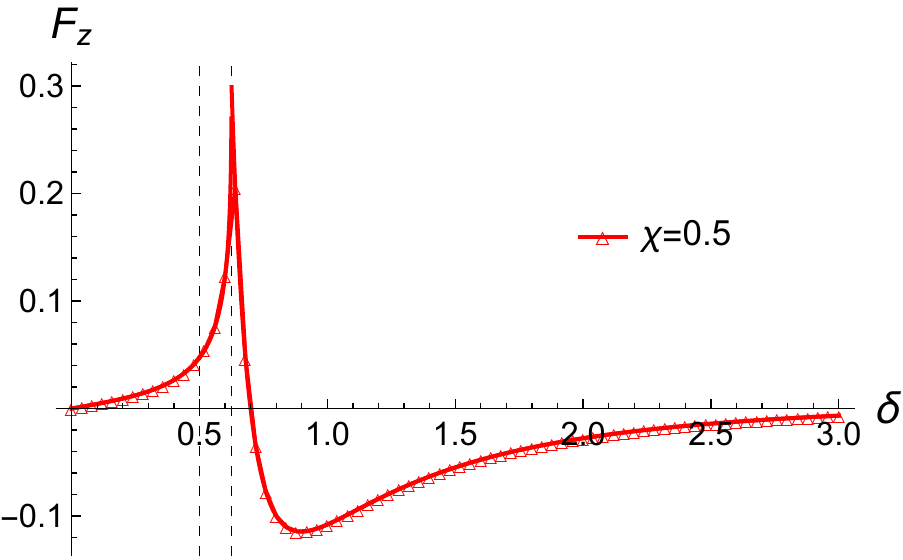}  \\ (c) \hspace{8cm} (d)
\caption{The upper panel show the functions $I _{\perp}$ (at left) and $I _{z}$ (at right) as a function of $\delta$ for a fixed value of $\chi$. The lower panel show the functions $F _{\perp}$ (at left) and $F _{z}$ (at right). }
\label{Functions}
\end{figure*}

Now we turn to the functions $F _{\alpha}$, with $\alpha = \perp , z$, as defined by Eqs. (\ref{Fperp}) and (\ref{Fz}). In this case, we apply a similar procedure to obtain analogous dimensionless expressions. Using the fact that $- \partial f _{s} ^{(0)} / \partial \mu = \partial f _{s} ^{(0)} / \partial E _{s \vec{k}} = - \delta ( \mu - E _{s \vec{k}} )$ and the definition of the functions (\ref{IntegrandFuncts}), we find that $F _{\alpha}$ can be written in the simple form
\begin{align}
F _{\alpha} (\mu ) &= - \frac{\hbar ^{2} v _{F} ^{2}}{ 4b } \frac{\partial}{\partial \mu} \sum _{s = \pm 1} s \int _{- \infty} ^{+ \infty} dk _{z} \int _{0} ^{\infty} dk _{\rho} \; k _{\rho} \, K  \, \mathcal{I} _{s \alpha} (k _{\rho} , k _{z}) \, \delta ( \mu - E _{s \vec{k}} ) . \label{SM-Fperp} 
\end{align}
The integral appearing in this expression can be evaluated in a similar fashion. As such, we omit the intermediate steps and move to the final result. In terms of the dimensionless variables $\delta$ and $\chi$, we obtain
\begin{align}
F _{\perp} (\delta , \chi ) &= - \frac{1}{4} \frac{\partial}{\partial \delta} \left\lbrace \Theta (\delta - \delta  _{ \mbox{\scriptsize vH}} ) \int _{\alpha _{--'}} ^{\alpha _{-+'}} + \Theta (\delta ) \Theta ( \delta  _{ \mbox{\scriptsize vH}} - \delta) \left( \int _{\alpha _{--'}} ^{\alpha _{+-'}} + \int _{\alpha _{++'}} ^{\alpha _{-+'}} \right)   \right\rbrace \frac{\alpha}{(\delta - \chi \alpha ) ^{2}} \left[ (\delta - \chi \alpha ) ^{2} - \frac{1}{4} (\alpha ^{2} - 1) ^{2} \right] d \alpha , \\ F _{z} (\delta , \chi ) &= - \frac{1}{4} \frac{\partial}{\partial \delta} \left\lbrace \Theta (\delta - \delta  _{ \mbox{\scriptsize vH}} ) \int _{\alpha _{--'}} ^{\alpha _{-+'}} + \Theta (\delta ) \Theta ( \delta  _{ \mbox{\scriptsize vH}} - \delta) \left( \int _{\alpha _{--'}} ^{\alpha _{+-'}} + \int _{\alpha _{++'}} ^{\alpha _{-+'}} \right)   \right\rbrace \frac{\alpha ^{2} - 1}{(\delta - \chi \alpha )} \left[ \chi + \frac{\alpha (\alpha ^{2} - 1)}{2 (\delta - \chi \alpha )} \right] d \alpha .
\end{align}

In Figs. \ref{Functions}(c) and \ref{Functions}(d) we plot  the functions $F _{\perp} (\delta , \chi )$ and $F _{z} (\delta , \chi )$, respectively, as a function of $\delta > 0$ and a given value of $\chi < 1$. In fact one can spot the identity $F _{\perp} (\delta , \chi ) + F _{z} (\delta , \chi ) = - I _{z} (\delta , \chi )$.
\end{widetext}

\section{Calculations of the nearly quantized case} \label{NearlyQ-App}

\begin{figure}[b]
	\centering
	\includegraphics[scale=0.25]{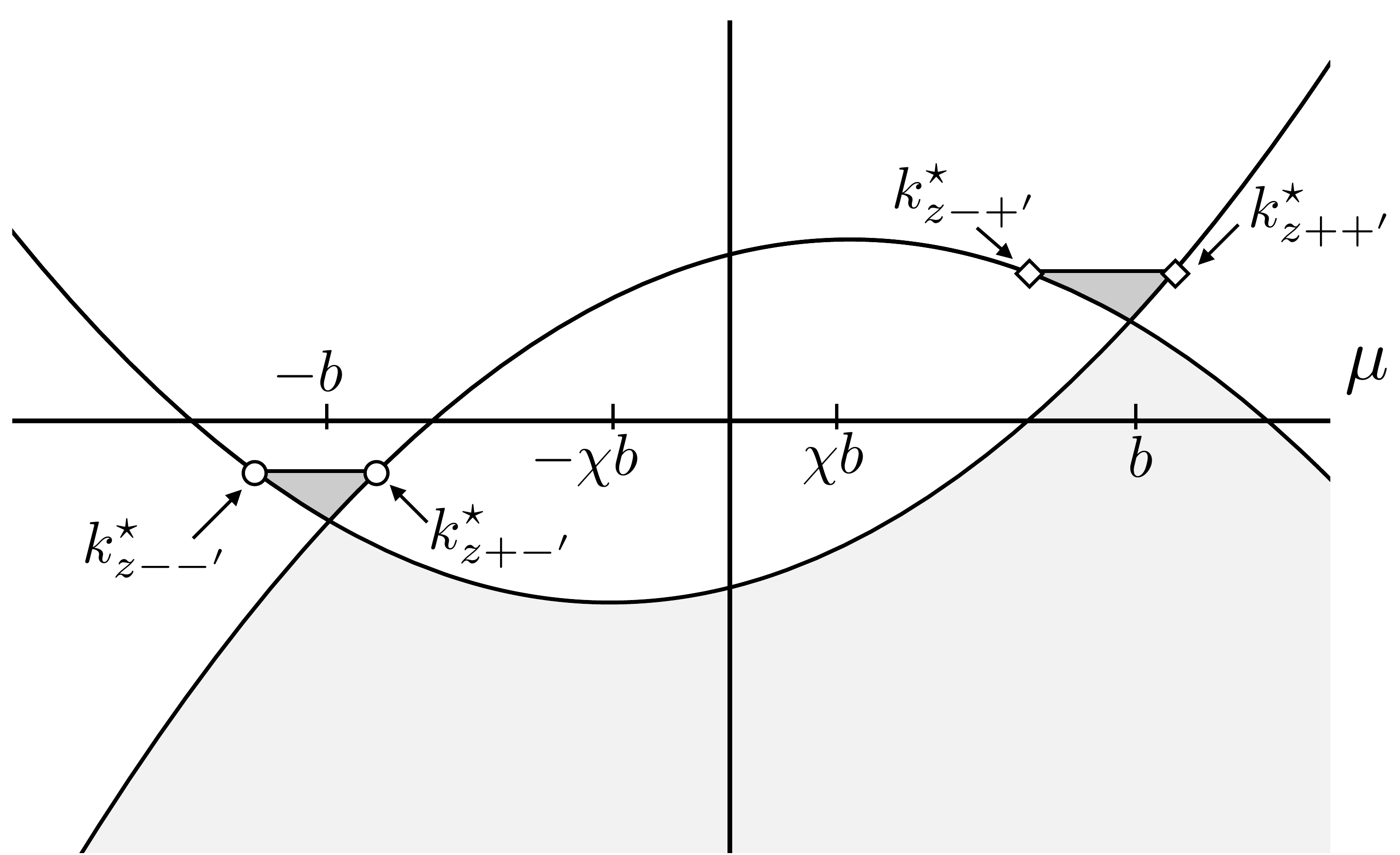}
	\caption{Schematic illustration of the integration regions of the functions $I_\alpha$ and $F_\alpha$ when the chemical potentials $\mu_L$ and $\mu_R$ are near to the left and right nodes, respectively. The points $k _{z \pm \pm ^{\prime}} ^{*}$ represent the crossings in the conduction band.
	}
	\label{IntRegionsquanta}
\end{figure}

As discussed in the main text, there is a particular case for which the conductivity tensor becomes nearly quantized. This occurs when the left and right chemical potentials are brought near to the corresponding left and right band touching points, as shown in the Fig. \ref{IntRegionsquanta}. In such case, only the conduction band contributes to the current and we can compute the coefficients $I _{\alpha}$ and $F _{\alpha}$ in a similar fashion. An explicit expression for the $I$'s is

\begin{figure*}
	\centering
	\includegraphics[scale=0.7]{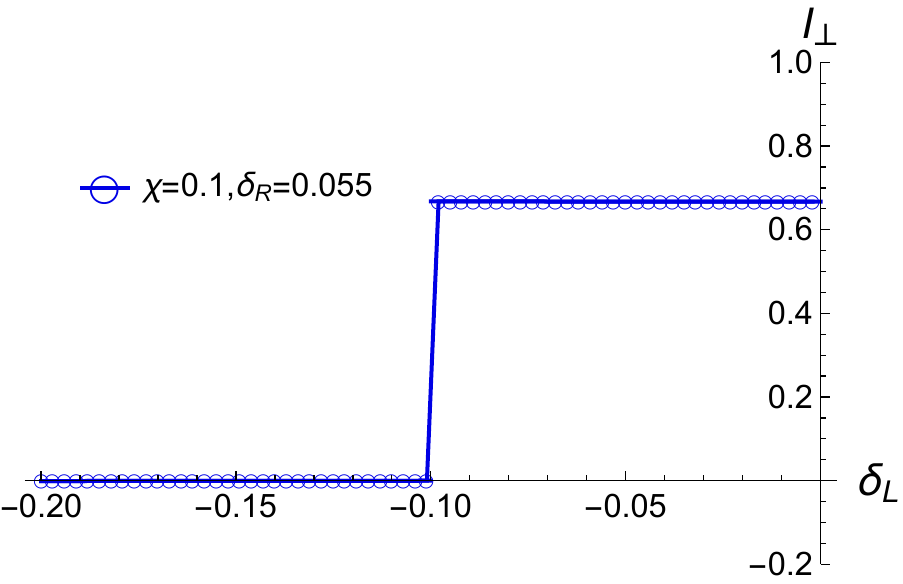}
	\hspace{0.1cm}
	\includegraphics[scale=0.7]{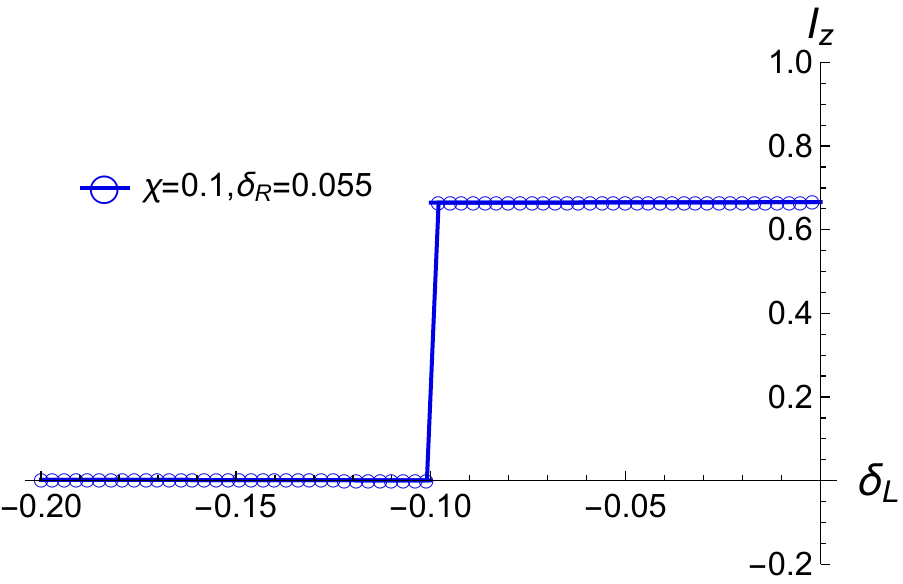} \\ (a) \hspace{8cm} (b) \\
	\includegraphics[scale=0.7]{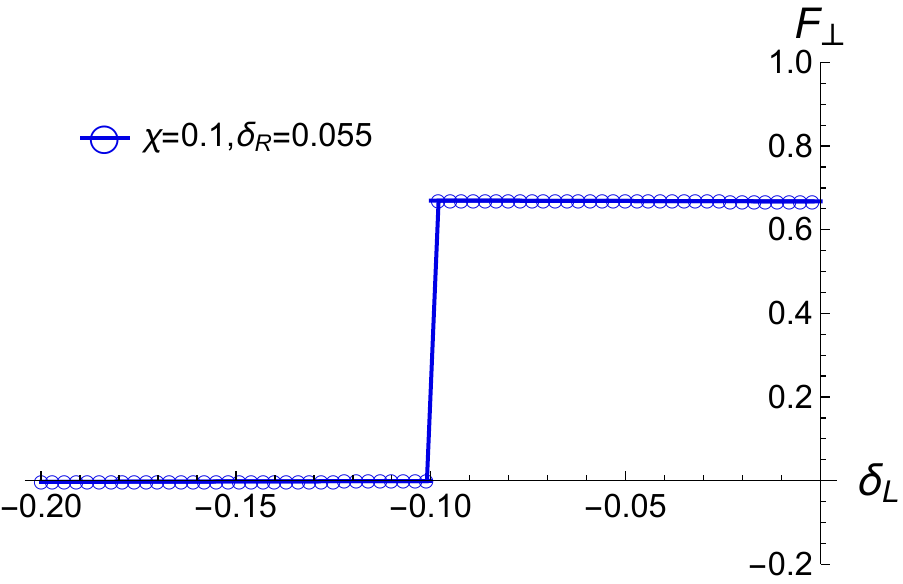}
	\hspace{0.1cm}
	\includegraphics[scale=0.7]{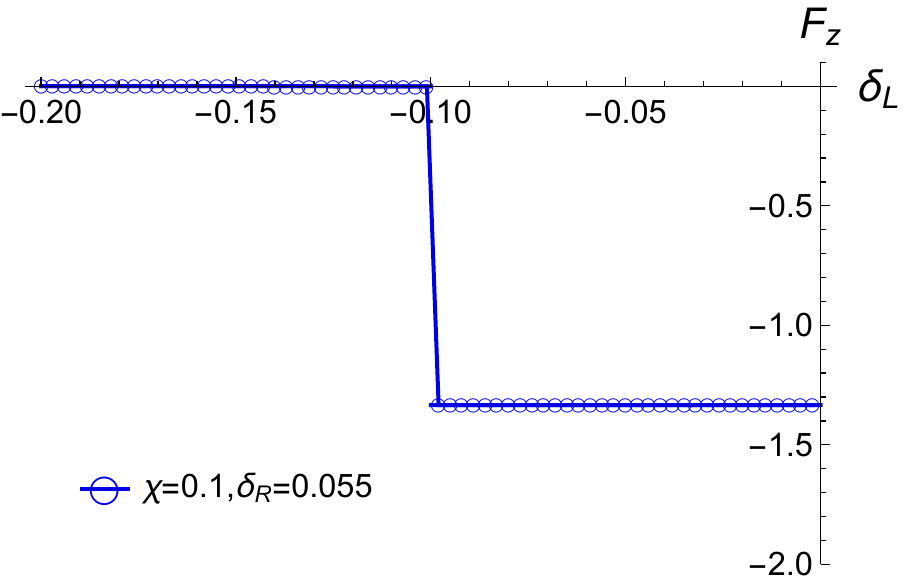}  \\ (c) \hspace{8cm} (d)
	\caption{The upper panel show the quantized functions $I _{\perp}$ (at left) and $I _{z}$ (at right) as a function of $\mu_L$ for a fixed value of $\chi$ and $\mu_R$. The lower panel show the functions $F _{\perp}$ (at left) and $F _{z}$ (at right). }
	\label{FunctionsQuanta}
\end{figure*}

\begin{widetext}
\begin{align}
I _{\alpha} ( \mu _{L} , \mu _{R} ) &= -  \int _{\boldsymbol{\mathcal{C}} _{L} (\mu _{L} )} \frac{\mu _{L}  - \hbar t k _{z}}{4b \hbar v _{F}}  \mathcal{I} _{1 \alpha} (k _{\rho} ^{\star} , k _{z}) dk _{z} - \int _{\boldsymbol{\mathcal{C}} _{R} (\mu _{R})} \frac{\mu _{R} - \hbar t k _{z}}{4b \hbar v _{F}}  \mathcal{I} _{1 \alpha} (k _{\rho} ^{\star} , k _{z}) dk _{z}
\end{align}
whose components can be expressed in terms of dimensionless variables as
\begin{align}
I _{\perp} (\delta _{L} , \delta _{R} , \chi ) &= - \int _{\alpha _{--'} ^{L}} ^{\alpha _{+-'} ^{L}} \frac{\alpha}{4 (\delta _{L} - \chi \alpha ) ^{3}} \left[ (\delta _{L} - \chi \alpha ) ^{2} - \frac{1}{4} (\alpha ^{2} - 1) ^{2} \right] d \alpha \notag \\ & \hspace{5cm}  -  \int _{\alpha _{++'} ^{R} } ^{\alpha _{-+'} ^{R}} \frac{\alpha}{4 (\delta _{R} - \chi \alpha ) ^{3}} \left[ (\delta _{R} - \chi \alpha ) ^{2} - \frac{1}{4} (\alpha ^{2} - 1) ^{2} \right] d \alpha 
\end{align}
and
\begin{align}
I _{z} (\delta _{L} , \delta _{R} , \chi ) = - \int _{\alpha _{--'} ^{L}} ^{\alpha _{+-'} ^{L}}  \frac{\alpha ^{2} - 1}{4 (\delta _{L} - \chi \alpha ) ^{2}} \left[ \chi + \frac{\alpha (\alpha ^{2} - 1)}{2 (\delta _{L} - \chi \alpha )} \right] d \alpha - \int _{\alpha _{++'} ^{R} } ^{\alpha _{-+'} ^{R}} \frac{\alpha ^{2} - 1}{4 (\delta _{R} - \chi \alpha ) ^{2}} \left[ \chi + \frac{\alpha (\alpha ^{2} - 1)}{2 (\delta _{R} - \chi \alpha )} \right] d \alpha , 
\end{align}
where $\alpha _{\pm \pm '} ^{L/R} = \pm \chi \pm ^{\prime} \sqrt{1 + \chi  ^{2} \mp 2 \delta  _{L/R}}$, and $\delta  _{L/R} = \mu _{L/R} / \hbar v _{F} b$.
 Similar expressions can be obtained for the $F$'s coefficients:
\begin{align}
F _{\perp} (\delta _{L} , \delta _{R} , \chi )  &= - \frac{\partial}{\partial \delta _{L}} \int _{\alpha _{--'} ^{L}} ^{\alpha _{+-'} ^{L}}  \frac{\alpha}{4 (\delta _{L} - \chi \alpha ) ^{2}} \left[ (\delta _{L} - \chi \alpha ) ^{2} - \frac{1}{4} (\alpha ^{2} - 1) ^{2} \right] d \alpha \notag \\ & \hspace{5.5cm} - \frac{\partial}{\partial \delta _{R}} \int _{\alpha _{++'} ^{R}} ^{\alpha _{-+'} ^{R}} \frac{\alpha}{4 (\delta _{R} - \chi \alpha ) ^{2}} \left[ (\delta _{R} - \chi \alpha ) ^{2} - \frac{1}{4} (\alpha ^{2} - 1) ^{2} \right] d \alpha , \\ F _{z} (\delta _{L} , \delta _{R} , \chi ) &= - \frac{\partial}{\partial \delta _{L}} \int _{\alpha _{--'} ^{L}} ^{\alpha _{+-'} ^{L}}  \frac{\alpha ^{2} - 1}{4 (\delta _{L} - \chi \alpha )} \left[ \chi + \frac{\alpha (\alpha ^{2} - 1)}{2 (\delta _{L} - \chi \alpha )} \right] d \alpha - \frac{\partial}{\partial \delta _{R}}  \int _{\alpha _{++'} ^{R}} ^{\alpha _{-+'} ^{R}} \frac{\alpha ^{2} - 1}{4 (\delta _{R} - \chi \alpha )} \left[ \chi + \frac{\alpha (\alpha ^{2} - 1)}{2 (\delta _{R} - \chi \alpha )} \right] d \alpha .
\end{align}
In Fig. \ref{FunctionsQuanta} we plot these functions and observe that they are nearly quantized as long as the chemical potentials approaches the band-touching points.
\end{widetext}

\section{Calculations of the fully quantized case} \label{FullyQ-App}

Substituting the velocity (\ref{VelChiral}) and Berry curvature (\ref{BerryCChiral}) into the definition (\ref{I-integral}) we obtain
\begin{align}
I _{\chi} &= \chi \frac{\hbar ^{3} v _{F} ^{3}}{6 \mu _{\chi} ^{2}} \int \delta ( \mu _{\chi} - \hbar v _{F} \vert \vec{k} - \chi \vec{b} \vert ) \, d ^{3} \vec{k} . \label{I-chiral}
\end{align}
To evaluate this integral we first express the Dirac delta function in terms of the roots of the argument by employing the formula (\ref{DiracDeltaProperty}). We obtain
\begin{align}
\delta ( \mu _{\chi} - \hbar v _{F} \vert \vec{k} - \chi \vec{b} \vert ) = \frac{\mu}{\hbar ^{2} v _{F} ^{2}} \frac{\delta \left( k _{\rho} - k _{\rho} ^{\ast} \right)}{ k _{\rho, +} ^{\ast}} \Theta \left( k _{\rho} ^{\ast \, 2} \right)
\end{align}
where $k _{\rho} ^{\ast} = \sqrt{(\mu _{\chi} / \hbar v _{F}) ^{2} - (k _{z} - \chi b ) ^{2}}$. Inserting this result into Eq. (\ref{I-chiral}) and introducing  a cutoff $\Lambda$ for the integration over $k _{z}$ we find
\begin{align}
I _{\chi} &=  \chi \frac{\pi \hbar v _{F}}{3 \mu _{\chi}} \int _{- \Lambda} ^{+ \Lambda}  \Theta \left(  \frac{\mu _{\chi}}{\hbar v _{F}} - \vert k _{z} - \chi b \vert \right) \; dk _{z} \notag \\ &= \chi \frac{\pi \hbar v _{F}}{3 \mu _{\chi}} \int _{- \frac{\mu _{\chi}}{\hbar v _{F}}} ^{+ \frac{\mu _{\chi}}{\hbar v _{F}}}  \; d \xi  = \frac{2 \pi}{3} \chi . \label{Fin-I-chiral}
\end{align}
We now turn to the integral  $F _{ \chi}$ defined in Eq. (\ref{F-integral}). In this case, it is convenient to express the integrand in terms of the energy $E _{\chi \vec{k}}$. The result is
\begin{align}
F _{\chi} = \chi \frac{\pi \hbar ^{3} v _{F} ^{3}}{3}  \int _{- \Lambda} ^{+ \Lambda} \int _{0} ^{\infty} \frac{1}{E _{\chi \vec{k}}} \; \frac{\partial ^{2} f _{\chi} ^{(0)}}{\partial E _{\chi \vec{k}} ^{2}} \, k _{\rho} d k _{\rho} d k _{z} .
\end{align}
So, by changing variable to the energy, i.e. $E _{\chi \vec{k}} d E _{ \chi \vec{k}}  = \hbar ^{2} v _{F} ^{2} k _{\rho} d k _{\rho}$, we obtain
\begin{align}
F _{\chi} &= \chi \frac{\pi \hbar v _{F}}{3}  \int _{- \Lambda} ^{+ \Lambda} d k _{z} \int _{\hbar v _{F} \vert k _{z} - \chi b \vert} ^{\infty} \frac{\partial ^{2} f _{\chi} ^{(0)}}{\partial E _{ \chi \vec{k}} ^{2}} \, d E _{\chi \vec{k}} \notag \\ &= \chi \frac{\pi \hbar v _{F}}{3}  \int _{- \Lambda} ^{+ \Lambda} d k _{z} \frac{\partial f _{ \chi} ^{(0)}}{\partial E _{ \chi \vec{k}} } \Bigg| _{ \hbar v _{F} \vert k _{z} - \chi b \vert} ^{\infty} .
\end{align}
Now, since $\frac{\partial f _{ \chi} ^{(0)}}{\partial E _{ \chi \vec{k}} } = - \delta (\mu _{\chi} - E _{ \chi \vec{k}})$ at zero temperature, after changing to the variable $\xi = k _{z} - \chi b$, the above integral reduces to
\begin{align}
F _{\chi} &=  \chi \frac{\pi \hbar v _{F}}{3}  \int _{- \Lambda - \chi b} ^{+ \Lambda - \chi b}  \delta \left( \mu _{\chi} - \hbar v _{F} \vert \xi \vert \right) d \xi . \label{F-integral}
\end{align}
To evaluate this integral we use the following property of the Dirac delta:
\begin{align}
\delta \left( \mu - \hbar v _{F} \vert \xi \vert \right) = \frac{1}{\hbar v _{F}} \left[ \delta (\xi - \xi ^{\ast} ) \Theta (\xi) + \delta (\xi + \xi ^{\ast} ) \Theta (- \xi) 
 \right] ,
\end{align}
where $\xi ^{\ast} = \mu / \hbar v _{F}$. Therefore, upon substitution of this result into Eq. (\ref{F-integral}) we finally obtain $F _{\chi} = ( 2 \pi / 3 ) \chi $.

\bibliography{ElectrochemicalTransportWSM}

\end{document}